\documentclass[manuscript]{acmart}
\usepackage{natbib}
\usepackage{colortbl}
\usepackage{multirow}
\usepackage{graphicx}
\usepackage{subfig}

\usepackage{comment}
\usepackage{longtable}
\usepackage{tipa}
\usepackage{soul}
\usepackage{pgfplots,stfloats}
\usepackage{multirow}

\AtBeginDocument{%
  \providecommand\BibTeX{{%
    \normalfont B\kern-0.5em{\scshape i\kern-0.25em b}\kern-0.8em\TeX}}}
\acmYear{2024}\copyrightyear{2024}
\setcopyright{rightsretained}
\acmDOI{10.1145/3630106.3658546}
\acmISBN{979-8-4007-0450-5/24/06}
\usepackage{soul}

\def\numofparticpants{15}
\def\numofquieres{61}
\def\numofhours{10}
\begin{document}

\title[Online Harm in Low-resourced Languages]{``I Searched for a Religious Song in Amharic and Got Sexual
Content Instead'': Investigating Online Harm in Low-Resourced Languages on YouTube.}


\author{Hellina Hailu Nigatu}
\email{hellina_nigatu@berkeley.edu}
\affiliation{%
  \institution{UC Berkeley}
  \country{USA}
}

\author{Inioluwa Deborah Raji}
\email{rajiinio@berkeley.edu}
\affiliation{%
  \institution{UC Berkeley}
  \country{USA}
}



\begin{abstract}
Online social media platforms such as YouTube have a wide, global reach. However, little is known about the experience of low-resourced language speakers on such platforms; especially in how they experience and navigate harmful content. To better understand this, we (1) conducted semi-structured interviews (n=15) and (2) analyzed search results (n=9313), recommendations (n=3336), channels (n=120) and comments (n=406) of policy-violating sexual content on YouTube focusing on the Amharic language. Our findings reveal that -- although Amharic-speaking YouTube users find the platform crucial for several aspects of their lives -- participants reported unplanned exposure to
policy-violating sexual content when searching for benign, popular queries. Furthermore, malicious content creators seem to 
exploit under-performing language technologies and content moderation to further target vulnerable groups of speakers, including migrant domestic workers, diaspora, and local Ethiopians. Overall, our study sheds light on how failures in low-resourced language technology may lead to exposure to harmful content and suggests implications for stakeholders in minimizing harm.
\textcolor{red}{\textbf{Content Warning}: This paper includes discussions of NSFW topics and harmful content (hate, abuse, sexual harassment, self-harm, misinformation). The authors do not support the creation or distribution of harmful content.}

\end{abstract}





\maketitle

\section{Introduction}

As the dominant video platform in several countries \cite{mozilla_youtube_2021, essential_nodate}, YouTube is a global phenomenon. One study \cite{atske_1_2019} indicates that around 67\% of all content on YouTube in 2019 was posted in a language other than English. In addition to serving as a source of entertainment, YouTube provides a way to generate income for many, including low-income families in developing countries~\cite{mozilla_youtube_2021}.
However, while online platforms such as YouTube bring benefits into the diverse contexts in which they operate, they also export their harms~\cite{mozilla_youtube_2021, Niu2022Nov, Thomas2022Apr, srba_auditing_2022}. 



%


Several government bodies around the world are trying to increase regulation over online content~\cite{esafty_principles_nodate, DepartmentforDigital2022Mar}. However, across many legal contexts, ``harmful content'' includes both legal and illegal material~\cite{keller2022lawful}, and so online platforms more reliably depend on their own ``Community Guidelines'' or other codes of conduct for moderating content on their platforms. Typically, platforms enforce their policies through a combination of human and automated content moderation pipelines~\cite{chandrasekharan_bag_2017, yin_detection_2009, roberts_behind_2021}. Despite these guardrails, users can still be exposed to harmful content due to, for example, a lack of proper policy enforcement \cite{Mackintosh2021Oct,Akinwotu2021Oct} or content moderation avoidance by content creators \cite{huertas-garcia_countering_2023}. Despite its global presence, and the platform's claim that ``Community Guidelines are enforced consistently across the globe,
regardless of where the content is uploaded''\cite{yt_community_guide}, previous work has already shown that ``Non-English speakers are hit the hardest'' when it comes to exposure to content the users ``regret''~\cite{mozilla_youtube_2021}. 
Likely, the under-performance of language technologies in low-resourced languages~\cite{goldfarb2023cross, nozza2021exposing, ojo2023good}\footnote{Following works from other disciplines~\cite{wingrove-haugland_not_2022, cummins_teaching_2017}, we exclusively use the term \emph{low-resourced} instead of \emph{low-resource} to emphasize that the low-resourcedness comes from \emph{choices} in design and development of language technologies that has left some languages---and by extension, the communities---behind, while these languages do not innately lack resource.}---i.e. languages that have limited digital data available~\cite{nekoto_participatory_2020, gu_universal_2018, zoph_transfer_2016}--- combined with the lack of linguistic competence in these languages amongst human moderators~\cite{TRGoogle}, make content moderation in these contexts especially challenging. 

While online content in low-resourced languages goes under-moderated, the experiences of users from these contexts are also under-studied\cite{wilkinson2022many, shahid_decolonizing_2023}. In this work, we study the experiences of low-resourced language speaking YouTube users. The YouTube transparency report~\cite{TRGoogle} indicates that policy-violating sexual content is amongst the most moderated types of policy violation on the platform. While not all online sexual content is harmful \cite{strohmayer_technologies_2019, ferris_tagging_2016}, online harm due to such content could occur in the form of non-consensual release of intimate videos and images \cite{gasso_victimization_2021}, unintentional exposure to sexual content \cite{woodhouse_regulating_2022}, and exposure of minors to sexual content \cite{livingstone_annual_2014}. Hence, we narrow our scope into studying \textit{policy-violating sexual content on YouTube in Amharic}, one of the local languages in Ethiopia with over 25 million native speakers worldwide~\cite{adugna_research_2022}.

We set out to answer the following research questions:

\begin{quote}
    
    \textbf{RQ1}: What are the experiences and coping mechanisms of users exposed to policy-violating sexual content on YouTube in a low-resourced language context? 
    
    \textbf{RQ2}: What are the characteristics of policy-violating sexual content on YouTube in Amharic? 


\end{quote}

To answer the above research questions, we conducted two studies. In \hyperref[study1]{\textit{Study 1}}, we conducted semi-structured interviews with low-resourced language speaking YouTube users to understand their experiences. In \hyperref[study2]{\textit{Study 2}}, we collected and analyzed search results, recommendations, and comments on policy-violating sexual videos in Amharic to understand the characteristics of the policy-violating content. 
To the best of our knowledge, this is the first study to directly and explicitly examine the experience of low-resourced language speakers with online harm on YouTube.



We describe our results in Sec. \ref{findings}, and discuss the implications of our findings from both studies in Sec. \ref{discussion}. 
We hope that our work informs social media platforms like YouTube in forming strategies for better enforcing existing policies to protect these marginalized communities.

\section{Related Works}

\subsection{Online Experiences and Harm} 
\label{previous_work_online_harm}

Global South communities are excluded in much of the design, development, and deployment of technologies~\cite{ansari_decolonizing_2019, dourish_ubicomps_2012, das_decolonization_2023, ali_brief_2016}. Often, such communities are left with technologies that are dysfunctional within their context or inadequate in addressing their needs. 
After surveying over 4000 participants from 14 countries, one study~\cite{schoenebeck_online_2023} found that the perception of online harm is highest in users outside the US. Another study~\cite {shahid_decolonizing_2023} found that Bangladesh Facebook users perceived content moderation systems as methods that propagate historical power relations and perpetuate colonial viewpoints. Previous work~\cite{wilkinson_many_2022} also found that Caribbean users had developed a perception of increased vulnerability and more severe harm than their US counterparts. 

Previous work has also shown the prevalence of harms associated specifically with sexual content. For instance, past work reveals how searching for benign terms such as ``black girls'' on Google~\cite{noble2018algorithms} or Google's Keyword planner~\cite{yin_google_2020} returns pornographic content. 
Further evidence reveals the toxicity in large, Internet-crawled datasets, including the existence of ``sexual abuse, rape, and non-consensual explicit images"~\cite{birhane_multimodal_2021}, indicating potential downstream impacts of online toxicity for machine learning model training and evaluation \cite{raji_ai_2021}.

\subsection{Challenges in Content Moderation} \label{previous_work_content_moderation}

Despite efforts by social media platforms, policies sometimes fall short in protecting users. 
Some online platform users employ content moderation avoidance strategies to circumvent platform policy restrictions~\cite{huertas-garcia_countering_2023, chancellor_thyghgapp_2016, moran_folk_2022, shahid_decolonizing_2023}. A common example of such techniques is \textit{lexical variation}, which involves modifying linguistic features (e.g phonetic) of constructs (e.g. words) in ways that do not alter their meaning, i.e. that are semantically equivalent \cite{Eisenstein2010Oct}. For example, lexical variants could occur via intentional misspellings or use of emojis and special characters \cite{chancellor_thyghgapp_2016, shahid_decolonizing_2023} (eg. ``seggs'' instead of ``sex''). While such tactics have been used to avoid over-moderation by non-malicious groups \cite{chancellor_thyghgapp_2016, shahid_decolonizing_2023} they have also been used to spread harmful content online \cite{huertas-garcia_countering_2023}.

Studies in multi-lingual Natural Language Processing (NLP) reveal that many of the automated functions of content moderation all fail to perform in the context of African languages, including Amharic -- this involves tasks such as sentiment analysis~\cite{goldfarb2023cross, talat2022you}, detection of hate speech ~\cite{nozza2021exposing, montariol2022multilingual,castillo2023analyzing} or social manipulation~\cite{haider_detecting_2023}, and even basic text classification tasks~\cite{ojo2023good}. Large Language Model content filters degrade in low-resourced language settings, effectively ``jail-breaking’’ such models to spew content that would otherwise be formally restricted~\cite{yong2023low}. Although there have been attempts to address this issue through novel benchmarks~\cite{muhammad2023semeval, muhammad2023afrisenti} and datasets~\cite{ayele20225js, yimam2019analysis, ayele2023exploring,jana_hypernymy_2022}, the current scope of interventions remains limited and fairly idiosyncratic -- for instance, the efforts in the Amharic language so far have only largely focused on deterring politically motivated hate speech on Twitter~\cite{ayele20225js}. 

The failure of content moderation schemes in Global South communities has, on several occasions, materialized into physical harm, including acts of violence and escalation of war in countries like Mayanmar~\cite{Mackintosh2021Oct,Akinwotu2021Oct}, Nigeria~\cite{adekoya2021information,mhaka2020social} and Ethiopia~\cite{ayele20225js}. Additionally, the impact of content moderation on Global South communities is not just in how moderation schemes fail to protect Global South users from online harm. Several articles shed light on the impact of Big Tech hiring content moderators from Global South countries and how these content moderators then experience harm in the form of low wages, lack of access to mental health support, and lack of protection by law~\cite{rowe_its_2023, noauthor_inside_2022, Musambi2023Jun, Hao2023Jul, wsj_big_2023, Kate_hidden_2023}. 

\subsection{YouTube Content Moderation Policies \& Practices} \label{yt_content_moderation}


On October 27, 2023, in compliance with the Digital Services Act (DSA), which came into effect November 16, 2022, Google released a Transparency Report,in conjunction with supplementary materials~\cite{harmful_YT,views_YT,viewsFAQ_YT}, on its products classified as a Very Large Online Search Engine (VLOSE) or a Very Large Online Platform (VLOP), including YouTube~\cite{TRGoogle}.
The report in particular provides concrete figures for Google's moderation efforts and resources during the period from August 28, 2023 to September 10, 2023. 
The report states, 
``\ldots{}Community Guidelines play an important role in maintaining a positive experience for everyone on our platforms \emph{no matter where they are in the world}''~\cite{TRGoogle}. Notably, these guidelines apply to not just the videos themselves, but ``all types of content on our platform, including videos, comments, links, and thumbnails''~\cite{harmful_YT}. YouTube claims that, in 2023, ``out of every 10,000 views on YouTube in Q2, only 9-10 came from violative content'', though it is unclear for which languages this evaluation was conducted~\cite{TRGoogle}. 



The company also explains its approach to detection. 91.3\% the content flagged for moderation is ``self-detected'' -- i.e. perceived through pro-active platform features and processes, rather than reported by users, or flagged through some other external means~\cite{TRGoogle}\footnote{Numbers calculated from Table 3.1.2 on ``the number of measures taken on violative content, broken down by service and detection method'', from Google Transparency Report~\cite{TRGoogle}}. They state that ``the overwhelming majority of violative content is detected by automated systems'', which is then re-directed to ``human reviewers''~\cite{TRGoogle}. These human content moderators ``evaluate whether (the content) violates our policies'' and ``...content assigned for their review may have been posted in several different languages. In some cases and where appropriate, translation tools may be used to assist in the review process and allow us to moderate content 24/7 and at scale''~\cite{harmful_YT}. YouTube reports that 89.2 \% of their human moderators operate in English. Of the non-English moderators, only 363 moderators (2.1\%) operate in a language other than the more highly resourced Spanish, Portuguese, French, German, or  Italian~\cite{TRGoogle}\footnote{Numbers calculated from Table 3.3.1 which "reflects the human resources evaluating content across the official EU Member State languages, for each service", from Google Transparency Report~\cite{TRGoogle}}. Once evaluated by human reviewers, ``we remove the content and use it to train our machines for better coverage in the future''~\cite{harmful_YT}.

Unless flagged for ``a clear educational, documentary, scientific or artistic (EDSA) purpose''~\cite{harmful_YT}, policy-violating content faces a set of articulated consequences -- most commonly ``Restrictions of the visibility of content'' (93.2\% of flagged content) 
~\cite{TRGoogle}, such as restricting viewership or disabling its recommendation-enabled reach\footnote{Numbers calculated from Table 3.1.3 on "Number of measures taken at Google’s own initiative, by service and type of restriction applied", from Google Transparency Report~\cite{TRGoogle}}. 
Over the reporting period, YouTube noted taking initiative on 6,320 cases related to ``Nudity / Sexual '' content -- significantly more than the number of actions taken on many of the other categories of online harm~\cite{TRGoogle}\footnote{Numbers taken from Table 3.1.1.h on "own initiative actions taken on YouTube, by type of illegal content or violation of terms and conditions", from Google Transparency Report~\cite{TRGoogle}}. This indicates that sexual content is amongst the most moderated types of policy violations on YouTube.


To understand YouTube's content moderation from the outside is much more difficult. There are several challenges in conducting audit studies on ever-changing, black-box platforms like YouTube. 
Prior works have used methods such as sock-puppet studies~\cite{srba_auditing_2022, sandvig_auditing_2014, bandy_problematic_2021} or browser plug-ins~ \cite{trackingexp_experience_2020, mozilla_youtube_2021} to circumvent these challenges. Other studies use qualitative methods focused on understanding the nuanced beliefs and behaviors of the users and creators who engage with the platforms~\cite{wu_agent_2019}. In our work, we combine both approaches, collecting platform data as well as conducting interview studies, to gain a wider perspective on the platform and user experiences.

\section{Methods}

In this section, we outline the methodological details for both of our studies. Each study went through and passed a formal Institutional Review Board (IRB) process at an accredited U.S. institution and was approved with IRB Protocol ID: 2022-11-15801.

\subsection{Study 1: Semi-Structured Interviews} \label{study1}
 Study 1 aims to answer \textbf{RQ1} through qualitative interviews discussing 
 the ~\emph{experiences} of Amharic-speaking YouTube users: how individuals become exposed to online harm, how they understand it, and what strategies, if any, they use to mitigate such harm. Further details can be found in Appendix \ref{guide_questions}.
 


\paragraph{Participant Recruitment}
We recruited participants through social media (Twitter, WhatsApp, Telegram), direct outreach to women's rights advocate groups in Ethiopia, and through the authors' personal networks. We used a screening survey to select participants. 
Following previous work~\cite{wu_agent_2019}, we exclusively recruited women as they are the population who would be most impacted by exposure to such content. Table~\ref{tab:particpants} presents participants' details.


\begin{table*}[]
    \centering
    \begin{tabular}{|p{1.5cm}|c|p{4.5cm}|c|p{2cm}|}
    \hline
         \textbf{Participant ID} &  \textbf{Profession} & \textbf{Languages} & \textbf{Location}  & \textbf{Years Using YouTube}\\
    \hline
         P0 &  Political Science & Amharic, Afaan Oromoo, Tigrinya, English & Ethiopia & 8 \\  \hline
         P1 &  Software Engineer & Amharic, English & Ethiopia & 6  \\ \hline
         P2 &  Management & Amharic, English & UAE & 10\\ \hline
         P3 &  Psychology & Amharic, Guragignya, Tigrinya, Arabic, English & Ethiopia & 9\\ \hline
         P4 &  Medical Student & Amharic, English, French & Ethiopia & 11\\ \hline
         P5 &  Communication and Media & Amharic, English & Ethiopia & 9 \\ \hline
         P6 &  Software Engineering Student & Amharic, English, French & Ethiopia & 8\\ \hline
         P7 &  Computer Scientist & Amharic, English & USA & 7\\ \hline
         P8 &  Natural Science Student & Amharic, English  & Ethiopia & 4 \\ \hline
         P9 &  Clinical Pharmacy Student & Amharic, Tigrinya, English & Ethiopia & 5 \\ \hline
         P10 & Medical Student & Amharic, English & Ethiopia & 8 \\ \hline
         P11 & Political Science Student & Amharic, Arabic, English & Ethiopia & 9 \\ \hline
        
         P12 & Computer Scientist & Amharic, Arabic, English & USA & 15 \\ \hline
         P13 & Natural Science Student  & Amharic, English & Ethiopia & 5  \\ \hline
         P14 & Management, Business Owner & Amharic, English  & Ethiopia & 7 \\ 

    \hline
    \end{tabular}
    \caption{\textbf{Table describing participant demographic for the semi-structured interview study (Study 2).} Participants reported \emph{\textbf{age in the range of 18-30}} and had a diverse set of professional backgrounds. Additionally, all participants spoke at least one Ethiopian language, with 3 participants speaking multiple Ethiopian languages. Our participant's minimum reported years of using YouTube was 5 while the maximum was 15. All participants consumed Amharic content on YouTube and 3 participants consumed content in Afan Oromo, Guragignya, or Tigrinya in addition to Amharic. All participants were women.}
    \label{tab:particpants}
\end{table*}

    \paragraph{Consent and Compensation} Participants signed a consent form approved by our IRB. Depending on their location, participants were compensated with a 25 USD gift card or a 1000 ETB mobile card top-up per hour of participation. 
    
    \paragraph{Procedure}
    We conducted semi-structured interview sessions remotely over Zoom, Google Meet, and through end-to-end encrypted calls on Telegram. Sessions lasted from 45-75 minutes. At the beginning of each session, we requested and received verbal consent (in addition to written consent) to record all sessions for subsequent analysis. Appendix \ref{guide_questions} presents our guiding questions. All interviews were conducted by the first author and were in a mix of Amharic and English, depending on participants' preferences.


\paragraph{Data Analysis}  We analyzed {\numofhours} hours of footage from {\numofparticpants} sessions and 9 pages of written notes. We used inductive thematic analysis \cite{braun_using_2006} to code the data. As the only author who speaks Amharic, the first author served as the interviewer, translator and transcriber as well as conducted direct open coding of each of the recordings with short, descriptive sentences using QualCoder \cite{qualcoder_2023}. All authors then discussed the open codes and synthesized higher-level themes, refining codes, and themes iteratively in weekly meetings -- resulting in a total of 936 unique open codes, grouped into 26 first-level themes and further grouped into 6 second-level themes (see Appendix \ref{apn:themes}).

\subsection{Study 2: Investigating Characteristics of Policy-Violating Sexual Content} \label{study2}

In order to characterize the nature of the policy-violating content identified in Study 1, we supplement our findings with 
an analysis of YouTube platform data to answer \textbf{RQ2}. 
Our language of study, Amharic, is primarily written in the Ge'ez script but sometimes also written in Latin characters online \cite{terefe_entropy_2017}---we conduct our analysis in both. 
Inspired by past work \cite{ribeiro_auditing_2020}, we minimized personalization by ~(1) not signing in,~(2) clearing all browser data before every query, and ~(3) running in a private, incognito window. To change locations, we used Virtual Private Network (VPN) services when using the YouTube platform, and use the coordinate parameter to specify the desired latitude and longitude when using the YouTube API. For the US, we collected data while physically being in the country. We collected all data using a Mozilla Firefox browser on a Linux machine during the period Dec 2022--April 2023.


\paragraph{Gathering Search Queries}
We assembled seed search queries from (1) the most common YouTube searches in Ethiopia \cite{top_searchs_ethiopia} and (2) common search queries from previous work \cite{nobel_algorithms_2018} auditing search engines, which we adapted to the Ethiopian context\footnote{For instance, we replaced `Black Women' with `Habesha Women'. ``Habesha'' is a term used to commonly refer to Semitic language-speaking communities in Ethiopia and Eritrea.}. We then ran an initial search, where we observed that sexual videos were returned for several of the top benign queries, including the names of famous children's TV shows. Additional benign tags identified on policy-violating content was added to our initial list of queries until we reached theoretical saturation \cite{strauss_discovery_2017}. 
Fig. \ref{fig:queries} presents our final query list \footnote{We added a screenshot of the table as we could not add Ge'ez characters with the currently allowed TAPS packages.}. The first author translated each English query into Amharic and rewrote Romanized Amharic words in Ge'ez; yielding a total of {\numofquieres} queries.

\paragraph{Search results}
We used the YouTube API \cite{youtube_api} to collect the top 50 search results for each query we shortlisted.
We collected the search results for each query in five different locations: Ethiopia, the United States (US), the United Kingdom (UK), the United Arab Emirates (UAE), and Saudi Arabia (SA), informed by common locations for local and diaspora Ethiopians. 
We collected a total of 9313 videos through YouTube search queries.


\paragraph{Video Recommendations}
To understand what kinds of policy-violating sexual videos are diffused by YouTube's recommendation algorithm, we open each policy-violating sexual video identified in the benign search phase, and captured all the recommended videos for each opened video.  
We conducted our data collection using Tracking Exposed \cite{tracking_2023}. 
Similarly to our search data collection scheme, we collected recommendation results in five different locations. 
We collected 3336 videos in total through YouTube recommendations.

\paragraph{Policy-violating Channels}
From the sexual videos returned in our data collection, we used the YouTube API to collect information about the channels that posted those videos. We collected information such as channel name, description, date of creation, location, number of views, number of subscribers, and number of videos. We collected data from 120 YouTube channels.

\paragraph{Comments}
From the channels in our data that posted the policy-violating sexual videos, we collected comments for three of the most popular videos from the channel with the most views. In total, we collected 406 comments, including threads, to understand the experience of YouTube users exposed to sexual videos.

\paragraph{Data Labeling and Analysis}

 For the videos collected from search and recommendation results, the first author opened each of the videos in the collected data and labeled the videos based on their title, thumbnail, audio, and video. For comments, we used inductive thematic analysis \cite{braun_using_2006}. The first author, as the only author who speaks Amharic, conducted line-by-line direct open coding of each of the comments in our collected data. Then, the first and second authors iterated over the codes and synthesized themes in frequent weekly meetings. Our analysis resulted in 163 unique codes which we iteratively grouped into 14 first-level themes which we further grouped into 3 second-level themes. For the channels, the first and second authors met weekly and categorized them based on the descriptions and list of videos. Throughout the labeling process, the first author provided translations for contents that were in Amharic. In total, the labeling and analysis lasted from Dec 2022 - July 2023. 

 \subsection{Limitations} In this study, we do not make claims about the ~\emph{prevalence} of policy-violating content in low-resourced languages. 
 Rather, in Study 2, we focus on attempting to understand the nature of the content we identify on the platform, in order to ground the characterization of experiences described in Study 1 with real YouTube platform data.
 We did not compare our results to any other language, as we are not interested in the \textit{degree} to which online harm occurs in low-resourced versus high-resourced languages. Instead, we focus exclusively on the chosen context (ie. of Amharic-speaking YouTube users) as a stand-alone contribution. This allows us to prioritize our primary research goal of understanding what online harm looks like in a non-Western community and how technological failures for low-resourced languages contribute to the speakers' exposure to online harm. Additionally, in Study 1, our participants are women who at the very least are attending undergraduate studies or have completed their Bachelor's degree. Additionally, Amharic is just one of over 80 languages spoken in Ethiopia. As such, we acknowledge our participant's sample is not representative of the realities of all women in the Ethiopian context and is biased towards educated women.  Due to these constraints, we acknowledge the limits of our findings toward broader generalizability. 

 \subsection{Positionality}
All authors of this study are of African descent, with a primary affiliation at a US-based university. The first author is a native Amharic speaker and 
also traveled to Ethiopia for part of the duration of the study, which eased technological and geographic barriers to communication, such as timezone differences and choice of communication platforms. 
We acknowledge our relative position of power and took steps to minimize the effects on our participants. In Appendix \ref{minimizing_harm}, we detail the steps we took to protect our participants from further harm, and how we employed the \textsc{HarmCheck} framework~\cite{derczynski_handling_2022} to reduce perpetuating harm to our participants, ourselves, and readers of our work.

\section{Findings} \label{findings}
In this section, we provide our findings from the two studies as they relate to our research questions. We summarize all the policy violations we observed from our two studies in Table  \ref{tab:policy_violations}.



\subsection{Low-resourced language speakers' experience deteriorates when using YouTube in their languages.} \label{experinces}

 All participants indicated using YouTube for educational and entertainment purposes. ``\textit{That is where we find tutorials and lecture videos for our course work}'' said P1. Further, some participants indicated using YouTube for news and politics (P0, P4, P11) and religious content (P3, P7, P10, P11). Some participants (P6, P7, P11) said if YouTube no longer existed, they would use different platforms for different content types, and that there is currently no single platform that could satisfy all their needs like YouTube. Several participants (P2, P5, P8, P10, P14) indicated TikTok might come close but criticized it for its short content length.

 
Our participants (P0, P3, P4, P7, P8, P11, P12, P14) reported that their search experience for Amharic was ``\textit{horrible}'' and ``\textit{requires a lot of scrolling}'', ``\textit{contains lots of unrelated content}'', and ``\textit{is full of click-baits and tabloids}''. P0 and P3 further state that they have similarly bad experiences when searching in Afaan Oromoo and Guragigna\footnote{Afaan Oromoo is written in Latin characters with the Qubee alphabet while Guragigna is primarily written with the Ge'ez alphabet.}, two other languages spoken in Ethiopia. Some participants (P5, P13, P14) reported having a bad search experience in Amharic unless ``\textit{you knew the exact title}'' and the ``\textit{exact spelling the content creators used}''. P5 further elaborates that search is fragile in Amharic, with something as small as a missing space offsetting the search experience. 
    \begin{quote}
         \textit{Amharic [words] in Latin [letters] have [many] different spellings \ldots{} sometimes you have to use 3 different [Latin] letters together for one Amharic letter. Usually what I do is use the suggestions on the auto-complete even if it is written in a way I would not normally write.}
                        ---P7
    \end{quote} 


P8 and P14 reported that searching for Amharic content in Latin script was worse than searching in Ge'ez while P2 reported having better experience when searching in Latin. All but one participant reported getting results in other languages such as Spanish, Russian, Hindi, Turkish, and other South Asian languages when searching for Amharic, Afaan Oromoo, or Tigrinya content, especially in Ge'ez scripts. 

    \begin{quote}
        \textit{When I search in Amharic, I get Amharic and languages I do not know. I am sometimes surprised and ask myself `I am still on YouTube, right?'}
            ---P1
    \end{quote}
    
From our interviews, we found that our participants reported being exposed to sexual videos when searching for ``\textit{Amharic Movies}'' (P1, P7), ``\textit{Habesha romance movies}'' (P11), when searching for Amharic music (P10), looking up Ethiopian artists (P5, P12), searching for benign general terms like ``\textit{Ethiopian girls}'' (P2), searching for literature work (P3), or searching for religious songs:    

\begin{quote}
        \textit{It was in the morning and I was about to pray \ldots{} I searched for a religious song in Amharic and got sexual content instead.}
                    ---P3
    \end{quote}

Participants (P3, P6) also reported encountering sexual videos as ``Up Next'' recommendations after watching non-sexual videos such as entertainment talk shows in Amharic. Some participants (P6, P9, P12) also reported getting sexual and violent videos on YouTube Shorts. P12 further elaborates that she saw these videos while ``\textit{mindlessly scrolling}''. Several participants (P1, P2, P4, P5, P8, P14) reported getting location-based recommendations which are usually related to politics and include hate speech or violent, graphic content. Additionally, participants had to use VPN services when YouTube was blocked in Ethiopia from February 2023 till July 2023 \cite{shutdown_ethiopia}. Some participants (P3, P8, P14) reported getting content in other languages although they did not report seeing a difference in the level or nature of harmful content. 

Participants (P0, P1, P4, P7, P9, P11, P12) reported seeing explicit words used in Amharic; ``\textit{in English it is censored \ldots{} in Amharic it is out in the air and there are no guardrails.}'' (P4). Participants (P0, P1) further elaborate that in some cases, content creators make efforts to blur explicit images while they have explicit Amharic writing. Others commented on how content creators use the conservative culture of Ethiopia to justify the release of such types of videos; by framing the videos as shaming women and putting women in their right place: 

\begin{quote}
    \textit{They use phrases like `she deserves to get a beating' and `the government should lock her up' to get clout and acceptance from the community for their content. What business does the government have over how an individual dresses? The culture is already conservative and content like this could translate to physical harm for these women.} ---P0
\end{quote}




\subsection{Once the Damage is Done: How low-resourced language speakers cope when policies fail.} \label{navigating_harms}


Our participants stated that they had previously reported videos with sexually explicit content (P0, P1, P5, P8, P12, P13, P14), religious and ethnic hate speech (P10, P11, P14), and graphic and violent content (P8). Two participants (P2, P10) said they had a successful reporting experience, where they got notifications the video was taken down. Meanwhile, others admitted that they either did not check after they reported (P9, P14), were told the video was not harmful (P6), or did not get feedback (P13, P9, P14). Participants (P0, P5) noted that they ``\textit{lost hope after repeated failed attempts}''. Some participants (P5, P6, P10, P14) also reported a sense of not feeling prioritized. One participant, whom we will avoid disclosing their participant ID to avoid any risk of de-identification, shared that they used to work for one of the big social media companies as a content moderator and that they had diminished hope and interest in using such platforms or relying on the reporting mechanisms. They further elaborated how online harm in low-resourced languages might have less priority: ``\textit{Our content might have small priority because \ldots{} it could be automated failures \ldots{} or lack of reporting by users.}'' Participants also expressed feeling like existing polices do not account for their cultural context. 
In fact, many (P3, P8, P11, P13, P14) inquired about who gets to define fairness or what is harmful:

\begin{quote}
   \textit{My country's and another country's definition of right and wrong is not the same. So I think they only take their own way of defining what is right. I do not think what I feel matters.}
            ---P14
\end{quote}


In addition to reporting, participants also indicated using the ``Do Not Recommend'' feature on YouTube (P1, P6, P9, P12) or just skipping the content (P3, P5, P7). ``\textit{\ldots{} even if you hover for a while, it will add it to your `watched' list and start recommending similar stuff \ldots{} I just skip things as fast as I can.}''(P7). One participant told us she used three different Google accounts for different content types:

\begin{quote}
    \textit{I created separate accounts: one for religious content, one for educational material, and one for entertainment and casual content. I didn't want to be hit with disturbing content when I was watching a religious sermon or looking at a lecture. My [channel] subscriptions are also different for each of the accounts. My search experience is still not ideal; sometimes I get Muslim content\ldots{} even though I exclusively search and consume Orthodox Christian content on that account \ldots{} but that is fine. } ---P10
\end{quote}

\subsection{Policy Violations In Search and Recommendation} \label{violations}
\textcolor{red}{CW: Includes discussion of rape.}
Using the snowball sampling approach described in Section \ref{study2}, we found that 19 out of 31 search queries returned at least 2 and at most 10 sexual videos in the top-50 results for Latin-based queries, with the highest number of sexual videos returned for the query ‘Doctor’ (see Fig. \ref{fig:latin_dist}). For Ge’ez-based queries, we found that 17 out of 31 queries returned at least 1 and at most 12 sexual videos in the top-50 search results, with the most number of sexual videos returned for a name of a children’s show, and the second highest number of sexual videos returned for the query ‘Habesha Doctor’ (see Fig. \ref{fig:geez_dist}) The impact of the identified policy violations is also influenced by other factors such as the placement of the video in the list of results and the nature of video thumbnails and titles. In Fig \ref{fig:docgeez}, we show examples of how searching for two benign queries results in sexual videos returned as the fifth and fourth search results respectively. Both results contain a neutral image of a woman, accompanied by provocative and explicit Amharic writing on the thumbnail and title. 
   
Next, we assess if policy-violating content is further diffused by YouTube's recommendation system. In the five locations we studied, we found that if a policy-violating sexual video is opened, a proportion of the recommended videos will also include policy-violating sexual content. In Appendix \ref{apn:screenshots}, we give visual examples: Fig.\ref{fig:verified} shows how, for a sexual video posted by a verified channel, all the recommendations are sexual videos from the same channel. 
Fig. \ref{fig:rec_vids}, shows examples of recommendations for sexual videos opened on YouTube. In some cases, these recommended videos included serious offenses, such as depictions of sexual violence. In one case, a migrant worker records the sexual abuse she experiences by her employer with a hidden camera. In another case, there is a narrative recording of sexual assault on public transportation, accompanied by explicit imagery. Recommendations also included videos of animals mating in parks or in zoos and sexual scenes cut from movies. Table \ref{tab:recdata} in Appendix \ref{apen:evidence} shows the percentage of policy-violating content in the recommendations across the five locations we studied.

\begin{table*}[]
    \centering
    \small
    \begin{tabular}{p{1.5cm}|p{6.8cm}|p{3.2cm}|p{3.2cm}}
    \hline
        \textbf{Policy} & \textbf{YouTube Violation Definition} &  \textbf{Violations from Study 1}&\textbf{Violations from Study 2}\\
    \hline
        Spam, deceptive practices, and scam  & Misleading Metadata or Thumbnails: Using the title, thumbnails, and description to
trick users into believing the content is something it is not &   Thumbnails that have nothing to do with the content (P4, P7, P10, P11)&Thumbnails that have nothing to do with the content (Search, Recommendation) \\
    \hline
        Playlists  & Playlists with thumbnails, titles, or descriptions that violate our community guidelines, such as those that are pornographic, or that consist of images that are intended to shock or disgust. &  Playlists with sexual narrations (P5)&Full sexual playlists (Search, Recommendation, Channels) \\
    \hline
        Child Safety & Sexualization of minors: Sexually explicit content featuring minors and content that sexually exploits minors. &  Sex tape of a minor; exposure to sexual content while a minor (P8, P2)&Sexual video involving a minor; comment from minor exposed to sexual video (Recommendation, Comments) \\
    \hline
        Thumbnails & Pornographic imagery. Imagery that depicts unwanted sexualization. Violent imagery that intends to shock or disgust. Graphic or disturbing imagery with blood or gore Vulgar or lewd language. A thumbnail that misleads viewers to think they’re about to view something that’s not in the video. &   Graphic and violent content; sexual imagery; explicit, vulgar language ( P0, P1, P2, P3, P6, P7, P8, P9 P13) &violent content, sexual imagery; explicit vulgar language (Search, Recommendation, Comment) \\
    \hline
        Nudity and Sexual Content  &  Explicit content meant to be sexually gratifying. Clips extracted from non-pornographic films, shows, or other content in order to isolate sexual content. Groping, kissing, public masturbation, “upskirting”, voyeurism, predatory exhibitionism, or any other content that depicts someone in a sexualized manner without their consent. Content that depicts sexual acts, behaviors, or sex toys that’s meant for sexual gratification &  Sex Tapes, Expose videos, Narrations, Rape, Sexual Acts, Sexual scenes from movies, Sexualization of women on streets, Nude leaks (P0, P1, P2, P3, P4, P5, P6, P7, P8, P9, P11, P10, P12, P13, P14)&Sex Tapes, Expose videos, Narrations, Rape, Sexual Acts, Phone Sex Recordings, Sexual scenes from movies, Nude leaks (Search, Recommendation) \\
    \hline
        Suicide, self-harm, and eating disorders & Videos showing the lead-up to a suicide, or suicide attempts and suicide rescue footage without sufficient context. &   Live suicide video (P7)&None \\
    \hline
        Vulgar Language  & Use of sexually explicit language or narratives. Use of excessive profanity in the content. Use of heavy profanity or sexually suggestive terms in the content’s title, thumbnail, or associated metadata. Use of excessive sexual sounds. &   Sexually explicit titles in Amharic; Vulgar words in thumbnails; Vulgar words in video content; Narrations of sexual acts (P0, P1, P3, P4, P7, P9, P11, P12)&sexually explicit titles in Amharic; vulgar words in thumbnails; vulgar words in video content; narrations of sexual acts, Explicit Comments (Search, Recommendation, Comments) \\
    \hline
        Harassment and Cyber-bullying  & Content featuring non-consensual sex acts, unwanted sexualization or anything that graphically sexualizes or degrades an individual. &  Harassment of women on streets; (P0, P1, P2, P6, P8)&Harassing Comments, Reactionary Videos Harassing Women (Recommendation, Comments) \\
    \hline
        Hate speech  &  Encourage violence and incite hatred against individuals or groups based on any of the attributes noted above.  &  Hate speech videos, songs with open call for violence, ethnic hate, religious hate  (P0, P4, P3, P10, P14)&Ethnic Slurs (Comments) \\
    \hline
        Violent or graphic content  & Inciting others to commit violent acts against individuals or a defined group of people. Footage or imagery showing bodily fluids, such as blood or vomit, with the intent to shock or disgust viewers. Footage of corpses with massive injuries, such as severed limbs. Graphic content that features animals and intends to shock or disgust. Violent physical sexual assaults (video, still imagery, or audio). &  Graphic videos of massacres, attacks; video showing bodily fluid (P0, P1, P3, P7, P8, P11, P13)&Rape caught on Camera (Search) \\
    \hline
        Misinform -ation  & Misattributed content. Content that may pose a serious risk of egregious harm by falsely claiming that old footage from a past event is from a current event. &  False news; Misattributed Content (P5, P6)&Misattributed Content (Search) \\
    \hline
    \end{tabular}
    \caption{\textcolor{red}{\textbf{CW: Discussion of sexual content, rape, self-harm and suicide, hate, abuse, and violence.}} \textbf{YouTube policy violations we observed from Study 1 and Study 2.} For Study 1, we present the ID of the participant who reported encountering online harm in low-resourced languages when using YouTube in Ethiopian languages. For Study 2, we indicate on which type of data collection and analysis we observed the content that violates the YouTube policy.}
    \label{tab:policy_violations}
\end{table*}

\subsection{On Strategies Employed by Policy-Violating Channels} \label{strategies}
 
In total, there were 131 unique channels identified from our search (n=30) and recommendation (n=110) data analysis that posted videos we labeled as containing policy-violating sexual content. 
25 out of 30 channels (83.33\%) identified through search exclusively posted Ethiopian content in Amharic while for the 53 of the 110 channels (48.18\%) identified through recommendations  posted Ethiopian content in Amharic.  We observed that from the channels that disclosed their location (n=90), around half of them(n=46) had their locations set to the US(n=35) or the UK(n=11), while none had their location set to Ethiopia. The year the channels were created ranged from 2006-2022 and of the 131 channels, 34 were ~\emph{verified}. Total number of views for the channels ranged from 558 views to 28 billion, while subscriber number ranged from 0 to 36 million. In Table \ref{tab:channels}, we show examples of policy-violating channels, along with their channel descriptions, and sample video titles to demonstrate the strategies used by the channels to promote their content. Narrowing in on the channels that post sexual videos in Amharic, we found that the content creators used the following content moderation avoidance strategies:

\begin{itemize}
    \item \textbf{Search Engine Optimization (SEO) Manipulation}: Placing hashtags with famous TV shows and celebrity names in the descriptions of their sexual videos. 
    \item \textbf{Presented credentials and appeals to authority}: Using channel names like ``Dr. [popular Ethiopian name]'', and having misleading descriptions that suggested they were providing ``health and lifestyle advice.'' 
    \item \textbf{Explicit Amharic Thumbnail Writings}: Posting content with thumbnails with explicit writing in Amharic next to stock images or neutral images of Ethiopian female celebrities.
    \item \textbf{Innocuous Visuals with Explicit Audio}: Posting videos with explicit, sexual Amharic audio while the accompanying visuals were something innocuous (e.g demo of how to use an online website-making service).
    \item \textbf{English Content as Disguise}: Manipulating the description box by putting benign text in English (e.g saying the video is about learning about different types of microbes) but explicitly sexual descriptions in Amharic.
    \item \textbf{Lexical variation}: Mixing phonetically identical Ge'ez and Latin letters when writing sexually explicit words.
    \item \textbf{Cross-Referencing}: Referring to similar channels as hashtags or in their channel descriptions.
    \item \textbf{Sharing Contact Information}: Posting phone numbers, usernames for other platforms, and email addresses. 
\end{itemize}




\begin{table*}[]
    \small
    \centering
    \begin{tabular}{p{.1\textwidth}|p{.2\textwidth}|p{.15\textwidth}|p{0.25\textwidth}|p{0.2\textwidth}}
    \toprule
        \textbf{Channel Info} & \textbf{Channel Description} & \textbf{Video Title} &\textbf{Video Description} & \textbf{Strategies used} \\
    \midrule
         \textbf{Name}: Dr.[NAME\_1] \textbf{Verified}:       Yes \textbf{Location}: United Kingdom       \textbf{Subscribers}: 169,000    \textbf{Video Type}: Narrations of sexual acts & Dr [NAME\_1] is Similar Channel to Dr [NAME\_3] and Dr [NAME\_2]. You can freely contact me with https://t.me/[TELEGRAM USERNAME] &  Dr [NAME\_1] \hl{How to make the lady [EXPLICIT WORD] repeatedly in one go.} dr [NAME\_1] insight habesha dr [NAME\_4] \#Ethiopia & Dr [NAME\_5] - Dr [NAME\_7] - Dr [NAME\_2] -[NAME] TUBE - DR [NAME\_3]  & Cross-Referencing; Providing contact information; Explicit Amharic writing in Thumbnail; Presented credentials and appeals to authority; SEO Manipulation \\
         \hline
         \textbf{Name}: Dr.[NAME\_6] \textbf{Verified}: No \textbf{Location}: UK    \textbf{Subscribers}: 52,400          \textbf{Video Type}: Narrations of sexual acts &  Thank you for subscribe! My name is dr [NAME\_6], in this channel you will be getting information about health care and skin routine tips and you [can] ask any [questions] about health care! If you [want] consulting you can call +[PHONE NUMBER] https://www.youtube.com /channel/ [CHANNEL ID] & Dr.[NAME\_6] \hl{If you want to [EXPLICIT WORD] her, touch her at these places[.] You will be amazed!} & Thank you for subscribing! \hl{things you need to know. for any questions related to [EXPLICIT WORD] you will find answers. All you need to do is subscribe. Become a member of this channel by subscribing.} Disclaimer: This Channel Does Not  Promote or encourage Any Illegal Activities, all contents provided by this Channel is meant for EDUCATIONAL PURPOSE Only. -  Telegram Address - @[TELEGRAM  USERNAME]. - Music from non-copyright music store \ldots{} \#dr[NAME\_1] \#dr[NAME\_3] \#ethiopiannews  & Presented credentials and appeals to authority; English as a Disguise; Cross-Referencing; Providing contact information; Innocuous Visuals with Explicit Audio; SEO Manipulation; Explicit Amharic writing in Thumbnail; \\
         \bottomrule
    \end{tabular}
    \caption{\textbf{\textcolor{red}{CW: contains discussion of sexual acts.} Examples of channels that post policy-violating sexual content in Amharic.} Basic information about the channel like name and number of subscribers is provided. We also list out which of the common strategies we identified the channel employs. All text translated from Amharic is highlighted in \hl{color}. For instance with channel Dr.[NAME\_6] We see that non-highlighted text (originally in English) mentions educational use and shows compliance with copyright while the highlighted text (Amharic) has explicit words. It also cross-references the first channel Dr.[NAME\_1] in the video description to feign legitimacy.}
    \label{tab:channels}
\end{table*}
 
\subsection{Disparate Experiences in the Comment Section} \label{comments}
Harmful content was not limited to the videos; we observed comments that spread hate, used slurs, used vulgar and sexually explicit language, and advocated for violence. In one thread, we observed an exchange where users over-sexualized a specific ethnicity and used ethnic slurs in the exchanges. There were also several instances of users employing the credentials presented by the content creators. For instance, several referred to the content creator as a ``doctor'', although there was no clear evidence of such a qualification beyond the channel's own description. Even those who disagreed with the content would say ``\textit{Doctors should not act this way}.'' or ``\textit{You should focus on giving medical advice.}'' Further, there was one comment that gave details about a real medical condition and asked for medical advice. We also found comments that support and further the demeaning of women, with some including borderline rape content, indicating violent acts, and disregard for consent. In Table \ref{tab:comments}, we give some samples of the comments we analyzed along with the themes and open codes associated with them. 

We observed several users indiscriminately share personal information such as age, marital status, and location. One commenter claimed to be 16 years old and expressed how the posted video impacted the way they think about intimacy. Of the comments that disclosed location information (7.06\%), about half (51.6\%) indicated being located in Middle Eastern countries, and some had indicators in their usernames and comments that they were migrant domestic workers (e.g using phrases migrant domestic workers have used to self-identify on online platforms). Some users would expose their location, and request others to pray for them to make it back to their country. In terms of marital status, we observed in the data that single people or people soon to be married would engage with the videos saying they ``\textit{will try it when I get married.}'' One comment had a phone number with a mix of words and numerals (e.g five5five0one0five \footnote{The example is a fictional phone number.  https://en.wikipedia.org/wiki/555\_(telephone\_number) }) and indicated ``\textit{women can contact me on [PLATFORM] if you are located in [CITY] and want abortion pills.}''.  

\begin{table*}[]
    \centering
    \small
    \begin{tabular}{p{0.15\textwidth}|p{0.2\textwidth}|p{0.2\textwidth}|p{0.4\textwidth}}
    \toprule
    \textbf{Second-Level Theme}  & \textbf{First-Level Theme} & \textbf{Example Open-Code} & \textbf{Example Comment}\\
    \midrule
\multirow{3}{0.15\textwidth}{Interactions among Commenters \\ ($N=$ 187, $C = 68$)} & Peer judgement & Using religion to scold user engagement & \hl{you are an impostor. You really listen to the Quran?}\\ \cline{2-4}
     & Sharing PII and personal details & Disclosing location & \hl{I will [EXPLICIT WORD] you. If you are in [CITY] let me give you my address.} \\ \cline{2-4}
    & Vulgarity & Threats of sexual violence & \hl{I want to [EXPLICIT WORD] you. I will coercively tie your hands and legs tightly and [EXPLICIT WORD] }\\ 
     
\hline
     \multirow{5}{0.15\textwidth}{Direct Engagement with Content \\($N=$ 187, $C = 65$)} & Emotive expression& Discussing arousal due to content exposure &\hl{Wow. I am like you too. But now I am [EXPLICIT WORD]}\\ \cline{2-4}
     & Advice giving & Sharing personal experiences & \hl{My husband really likes it when I touch his [EXPLICIT WORD]} \\ \cline{2-4}
     & Expressing encouraging, positive sentiments & Expressing gratitude & \hl{Thank you, doctor, for the wisdom you share with us}\\ \cline{2-4}
     & Engaging in medical discourse & Sharing health information & \hl{below my belly button, my uterus.}\\ \cline{2-4}
     & Expressing positionality & Explaining personal context or mode of engagement with the content& \hl{I am now married. I am listening to you attentively for my husband's sake.}\\ \cline{2-4}
     & Asking for more & Asking the content creator to add visuals & \hl{Why don't you show us with video?} \\ \cline{2-4}

\hline     

     \multirow{4}{0.15\textwidth}{Opinion about Content \\($N=$ 75, $C = 30$)} & Expressing concerns & Arguing that content is inappropriate & \hl{Please use words that are acceptable in Ethiopian culture. Please don't ever use vulgar words. Seems like you are forgetting this is social media.}\\ \cline{2-4} 
     & Questioning channel legitimacy & Questioning the creator's medical knowledge and credentials & \hl{are you really a doctor? first, treat yourself. There are plenty of medical topics that need discussion. You are just doing this to get more subscribers.}\\ \cline{2-4}
     & Expressing approval & Defending content creators from critique & \hl{:laughing: this is something God created. Where is the inappropriateness?}\\ \cline{2-4}
     & Fear of societal consequences & Concerns for impact to younger generation & I'll report your channel for inappropriate sexual content, there are young kids exposed to your filth \\  

     
\bottomrule
    \end{tabular}
    \caption{\textbf{\textcolor{red}{CW: contains discussion of sexual violence.} Analysis of Comments on Policy-Violating Sexual Content.} Here, we present the number of codes assigned to each theme ($C$) and the number of comments coded with each theme ($N$). In total, we analyzed 406 comments -- 371 were legible. Legible comments were open-coded with a total of 163 unique codes, grouped into 14 first-level themes, and 3 second-level themes. Examples codes are provided in the table. Note that some comments have multiple codes.}
    \label{tab:comments}
\end{table*}

Some comments (5.17\%) explicitly opposed the content. Critical commenters noted, ``Such content is against our cultural beliefs'', and ``This is religiously unacceptable'', expressing fear that 
``There are young kids exposed to this type of content''. 
One comment opposing a video that had advice for violent sexual acts indicated how the suggestions are damaging to women’s health:

\begin{quote}
    \textit{What type of demonized lesson are you giving to men? You have never experienced what it is like to be a woman and yet you are talking as if you are an expert... You are going to ruin people's marriages with this content.} ---Comment from a YouTube user, translated from Amharic
\end{quote}


\section{Discussion} \label{discussion}



Our findings indicate that Amharic-speaking users report being regularly exposed to policy-violating sexual content on the platform (Sec. \ref{experinces} and  Sec. \ref{violations}). This could be due to malicious content creators relying on the under-performance of content moderation in low-resource languages (Sec. \ref{strategies}), or a set of other related factors.  Regardless, our work suggests that while low-resourced language speakers find the platform genuinely enriching, 
they feel as though existing policies are being inadequately enforced. 
As a result, these users develop their own mitigation strategies beyond platform reporting in order to minimize the personal impact of online harms (Sec. \ref{navigating_harms}). 
Our work shows how, even when policies exist, policy-violating content in low-resourced languages harms users due to (1) lack of proper enforcement by platforms and (2) due to under-performance of language technologies for low-resourced settings (Sec. \ref{previous_work_content_moderation}). While recent efforts by community-led research groups~\cite{litre_participatory_2022, meyer_bibletts_2022, dossou_afrolm_2022, adelani_few_2022} have contributed works to move the NLP community in a less Anglo-centric~\cite{noauthor_benderrule_2019} direction, we show how the exclusion of low-resourced languages in mainstream NLP, along with exclusion in policy enforcement by platforms, has real-world, immediate downstream harm.

\paragraph{Recommendations for Social Media Platforms with A Global Reach}

 Our results indicate weak moderation efforts on these platforms in the low-resourced language context. Failure to detect harmful content in low-resourced languages (Sec. \ref{violations} ) allows malicious content creators to surpass guardrails put in place by the platforms and leaves their users vulnerable to harmful content (Sec. \ref{strategies}). Additionally, search and recommendation features of YouTube further the reach of policy-violating content on the platform (Sec. \ref{violations}). As Fig. \ref{fig:verified} shows, a channel that posts policy-violating content in Amharic is (1) verified by the platform and (2) recommendations to videos from the channel are exclusively other policy-violating videos from the same channel.  Furthermore, users feel a sense of not being prioritized and their cultural context not being accounted for in content moderation (Sec. \ref{experinces}). Hence, (1) even when policies exist, they are not properly enforced in low-resourced language contexts and (2) users feel moderation strategies and current policies are not reflective of the diverse cultural contexts and realities of global users. Platforms should be aware of and actively consider their limitations within the low-resourced language setting, and potentially explore new directions for culturally aware and context-specific moderation strategies. This echoes previous findings \cite{schoenebeck_online_2023} advocating for the inclusion of diverse perspectives in content moderation.  While warning against the \emph{over-penalization} of marginalized communities~\cite{haimson_disproportionate_2021}, we argue that the effectiveness of online platform policy enforcement cannot be defined independently of cultural appropriateness or the cautious identification of policy violators. 

\paragraph{Recommendations for NGOs focused on Protecting Marginalized Groups}
By exclusively studying the experiences of women who speak low-resourced languages, our findings show how these marginalized groups can be exposed to and genuinely harmed by policy-violating sexual content on YouTube. We observed that vulnerable groups such as migrant domestic workers, who may not have their own access to healthcare services~\cite{noauthor_lebanon_2019}, might instead engage with videos by malicious content creators claiming to be medical doctors (Sec. \ref{strategies}). This puts them at risk of, for example, exposing their PII and medical information in the comments (Sec. \ref{comments}). Our findings can help support NGOs working with marginalized communities (ie. women, children, migrant workers) in providing training regimes or awareness campaigns for effective online navigation tactics. 

\paragraph{Recommendations for Government Bodies Hoping to Protect their Citizens.}

Content moderation failures have exposed citizens of Global South countries to physical and psychological harm--both as users of the platforms and employees for content moderation (Sec. \ref{previous_work_content_moderation}). Efforts by individual governments \cite{oluyemi2023Aug} and collective agencies like the AU \cite{au_pioneering_2023, au_african_2023, nepad_artificial_2023} should emphasize the culturally and contextually appropriate protection of their citizens from online harms. Government bodies could, for instance, set up oversight agencies that require platforms to demonstrate and disclose how their policies account for the languages of the countries in which they operate. 

\section{Conclusion}

Our paper investigates the experience of Amharic-speaking users with policy-violating sexual content on YouTube. Our findings shed light on a small number of individual experiences within one of many low-resourced language speaking communities -- in fact, Amharic is just one of the many low-resourced languages spoken in Ethiopia alone. More work is needed to further investigate the role of language in characterizing online experiences in other settings -- i.e. multiple geographic regions, more languages, more dialects, involving more extensive platform data collection on a wider array of online harms. While conclusions may emerge that are common to low-resourced languages in general, the intersections of culture, language, and identity unique to each community or even each individual mean that we should be hesitant to generalize findings too quickly, and should instead prioritize in-depth, particular research on specific groups and individuals over broader theories. Overall, we hope our study will help inform policy-making, digital literacy research, and technological design on online platforms to properly protect and serve users operating in low-resourced language contexts. 



\begin{acks}
Hellina Hailu Nigatuis a SIGHPC Computational and Data Science Fellow. Inioluwa Deborah Raji is a Senior Fellow at Mozilla. We would like to thank members and friends of PLAIT lab for their feedback on this work. Additionally, we thank Daricia Wilkinson and Niloufar Salehi for their consistent feedback and guidance on this work. We thank Abeba Birhane for reviewing and giving feedback on versions of the manuscript. We thank Liza Gak and Sijia Xiao for the early conversations regarding this work. We also want to thank our colleagues from Addis Powerhouse. Finally, to our participants, we say Enamesegnalen.
\end{acks}

\bibliography{references, custom, sample-base} 

\appendix

\section{Guiding Interview Questions} \label{guide_questions}
In this section, we present the guiding questions we used during our interviews for Study 1. Since it was a semi-structured interview, we used those questions as a guide and used participants' responses to probe further into a particular theme. We arrived at this set of questions after continuous feedback from colleagues, specifically on making the questions open-ended to avoid leading interviewees to an answer. 

\textbf{Demographic Questions}

\begin{itemize}
    \item What is your age range?
    \item What languages do you speak?
    \item Are you a student? Professional?   What field? 
\end{itemize}                  

\textbf{About use of YouTube}

\begin{itemize}
    \item What do you use YouTube for? 
    \item What kind of content do you consume on YouTube?
    \item What features of YouTube do you use?
    
    \item Do you have your own channel?
    \item Do you post on your own channel?
 
    \item Do you comment on videos?
    \item Do you like videos?
    \item Do you subscribe to channels? What factors do you consider when you decide to subscribe?
    \item What kinds of channels do you subscribe to?
\end{itemize}

\textbf{YouTube as a platform}
\begin{itemize}
    \item In your opinion, what is a good substitute for YouTube?
    \item What kinds of trends do you observe over the years as you use YouTube?
    \item How would you describe your YouTube homepage? Do you think everyone's home page is the same?
    \item How do you think YouTube works?
\end{itemize}

\textbf{Reporting}

\begin{itemize}
    \item What type of content have you reported? 
    \item What do you think is YouTube’s policy for reporting?
    \item How do you go about reporting?
    \item Why have you not reported yet?
\end{itemize}

\textbf{Language related questions}

\begin{itemize}
    \item In what language do you use YouTube? Why?
    \item In what language do you comment on YouTube? Why?
    \item In what language do you consume YouTube content? Why? 
\end{itemize}

\textbf{About Harmful content}
\begin{itemize}
    \item What kinds of harmful content have you experienced on YouTube?
    \item How would you describe the way you reach these types of content? Through search, recommendation?
    \item What does recommendation mean?
    \item How do you think YouTube recommendations work?
    \item Have you ever reported videos on YouTube? How?
    \item Have you seen a difference in the type of harmful content you get depending on the language you are using?
\end{itemize}

\section{Minimizing Harm} \label{minimizing_harm}

We iteratively shared drafts of the paper with participants who were concerned with de-identification and refrained from sharing details about participants per their comfort level. While this limited the demographic information we can share, our participants' safety and concern are our top priority. Additionally, we minimized potential biases in our study by clarifying to participants they can refuse to answer any question and only delve into details depending on their comfort level. One participant, after stating she had been exposed to sexual videos on YouTube and giving a general description, declined to go into specific detail about the types of sexual videos she encountered. In this case, we instead focused on how she was exposed to such videos and what strategies she used to mitigate harm. We make it clear that declining to answer any question will not affect their compensation in our screening form, consent form, and during the interview sessions. We also acknowledge the privilege we had in stepping away from the data when the mental load was too much. 


\subsection{Documentation of harmful text presentation the HarmCheck framework.} \label{harmchek}
In this section, we use \textsc{HarmCheck}\cite{derczynski_handling_2022} as a framework to scrutinize how we presented the harmful content in our research.  

\textbf{Risk of harm protocol} There is a risk of harm to data subjects whose images and names have been used in the distribution of sexual content and to subjects whose identities have been described using slurs in the data. There is a risk of harm to the researchers who have to label and analyze the data. The study included frequent check-ins, 10-20 minute conversations over the impact of the data labeling process on the researcher's mental health, and support and understanding for taking breaks from the project. There is a risk of exposing harm to those sharing work and home environments. The research was conducted either in isolation (outside of the lab) or with a content warning sent to lab mates. Additionally, we restricted ourselves to only processing text data in Amharic in the lab (with content warnings still going out) since no other lab mate could read or understand the language. There is also a risk of harm to the reader from exposure to explicit, distressing, and/or offensive images. We try to minimize this risk by providing content warnings and blurring images. 

\textbf{Preview} A content warning is included directly after the abstract and is highlighted in red. It is placed before any in-depth discussion of harmful content and describes the nature of the content. Captions in tables and figures also include content warnings to warn against the type of harmful content included.


\textbf{Distance} We avoid propagating harm by minimizing the examples we provide and replacing them with identifiers. In cases where interview subjects describe the content they were exposed to in detail, we avoid further propagation of harm by reducing direct quotes in those scenarios and instead giving general descriptions.


\textbf{Disclaim} In order to clearly indicate that the images and text come from the platform, we used screenshots (which we then blurred) that show the platform's interface. 


\textbf{Respect} We blurred images to protect subject identity. We also removed all personally identifiable information (including location) from interviews and from data we collected. Additionally, for interview subjects concerned with their data protection, we shared the anonymized table (Table \ref{tab:particpants}) and made sure they were okay with the presentation. We also do not release any data publicly. We avoid presenting data that shows groups being described with pejorative terms and instead describe the general phenomenon or use identifiers. We also use placeholders for celebrity names and TV shows we used in our queries out of respect for individuals and to protect individuals and groups from further harm.


\section{Screenshots} \label{apn:screenshots}
\begin{figure*}
          \centering
          \subfloat[Screenshot showing the results for searching for Doctor in Ge'ez script.]{\includegraphics[width=0.4\textwidth]{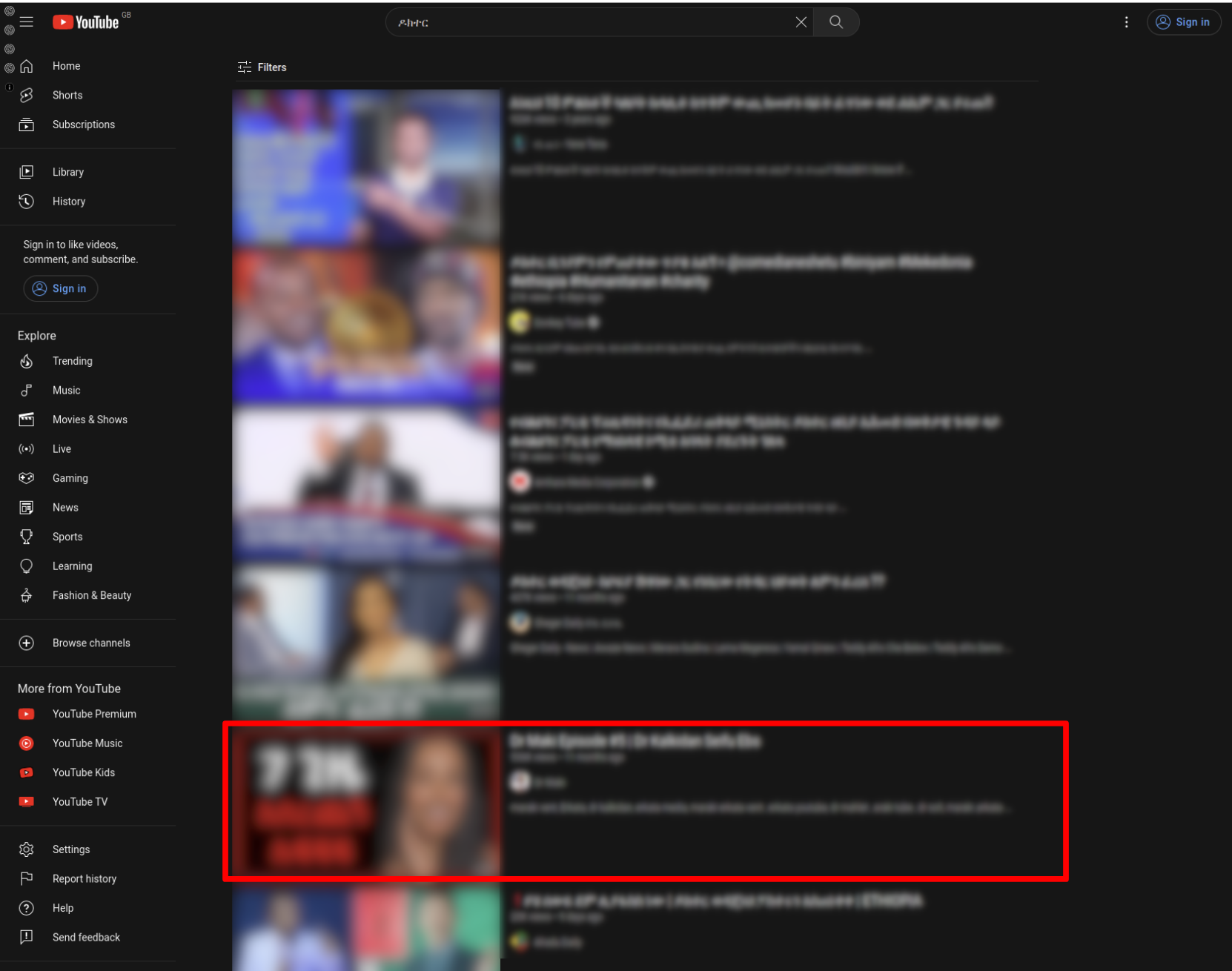} \label{fig:docgeez}} 
          \hfill
          \subfloat[Screenshot showing the results for searching for a famous TV show in Latin script.]{\includegraphics[width=0.4\textwidth]{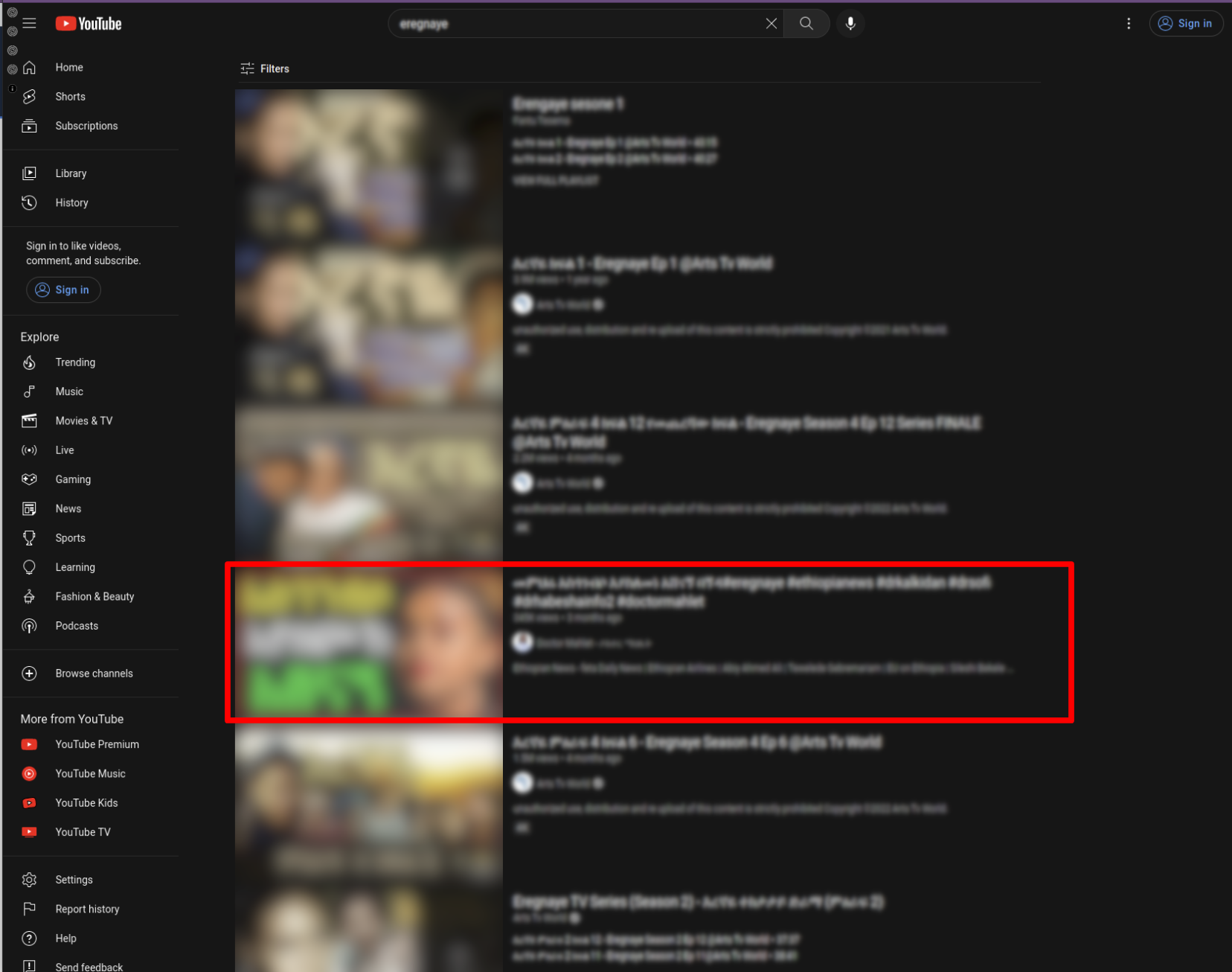} \label{fig:tvshow}} 
            \hfill
          
          \caption{\textcolor{red}{\textbf{CW: Discussion of sexual content.}} In Fig.\label{fig:docgeez} we present a screenshot for the search results for the query ``Doctor'' written in Ge'ez script. The first four results are from a medical YouTube channel by Ethiopian doctors, an entertainment talk show, a news channel, and a talk show featuring a psychiatric doctor. Then, the fifth result (highlighted in red box) is a sexual video with a picture of an Ethiopian girl and explicit sexual writing on the thumbnail. The title says `He made me [EXPLICIT WORD] 7 times'. The video is from a channel that has a name that starts with ``Dr.'' Similarly, in Fig. \ref{fig:tvshow}, the first three results for a famous TV show are episodes of the TV show, and the fourth result (highlighted in red box), is a sexual video with similar characteristics as the one in Fig. \ref{fig:docgeez}. }
      \label{fig:search_benign}
      \vspace*{-13pt}
      
        \end{figure*}

\begin{figure}
        \centering
        \includegraphics[width=0.4\textwidth]{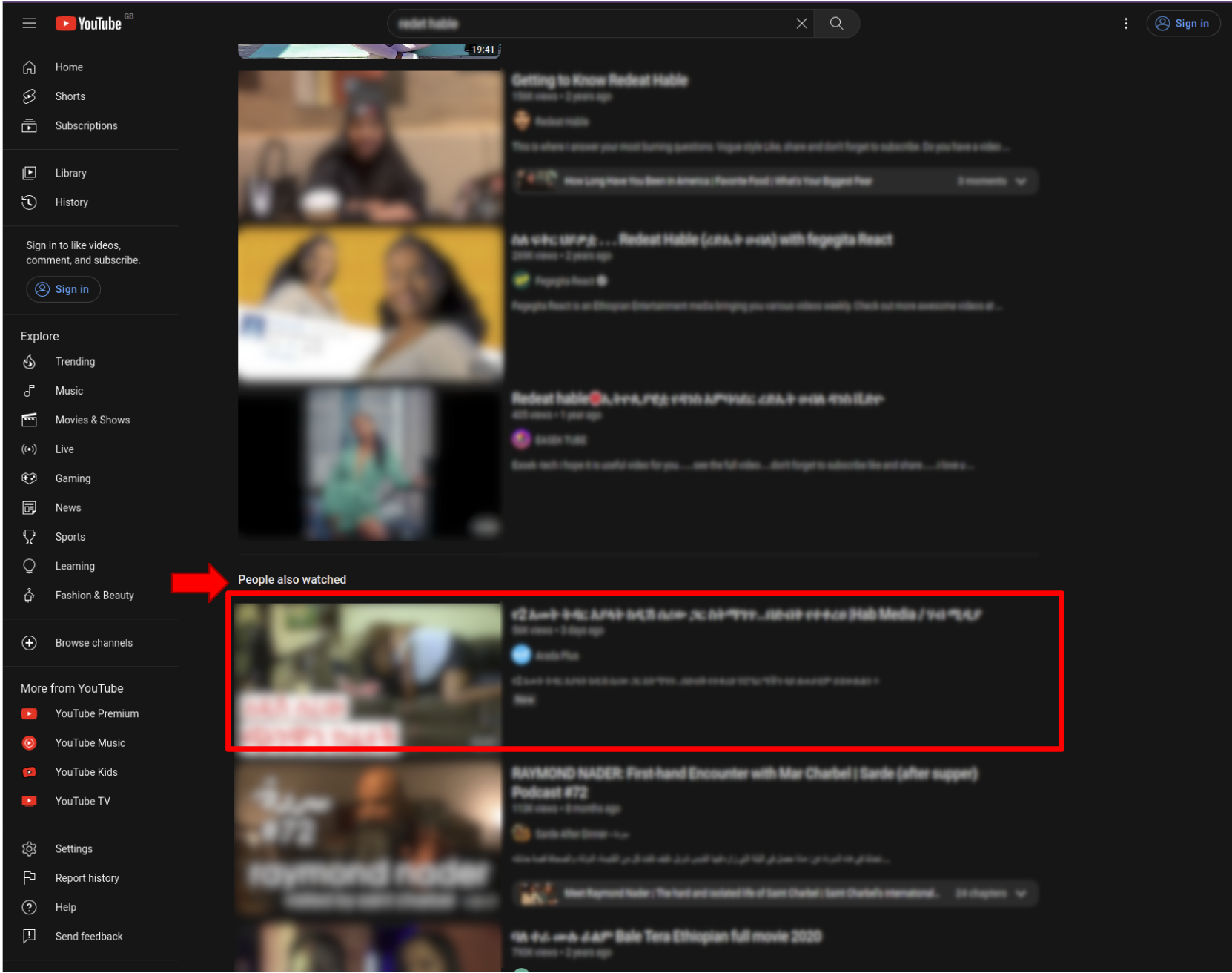}
        \caption{\textcolor{red}{\textbf{CW: Discussion of sexual content.}} Scrolling down the search results for a famous Ethiopian celebrity. Sexual videos are not limited to direct responses to search queries. Here, we found a sexual video that has two people engaged in a sexual act on a sofa with the title indicating a person cheating on her husband with a satellite dish maintenance person in the ``People also watched" section.}
        \label{fig:red}
    \end{figure}

\begin{figure}
        \centering
        \includegraphics[width=0.5\textwidth]{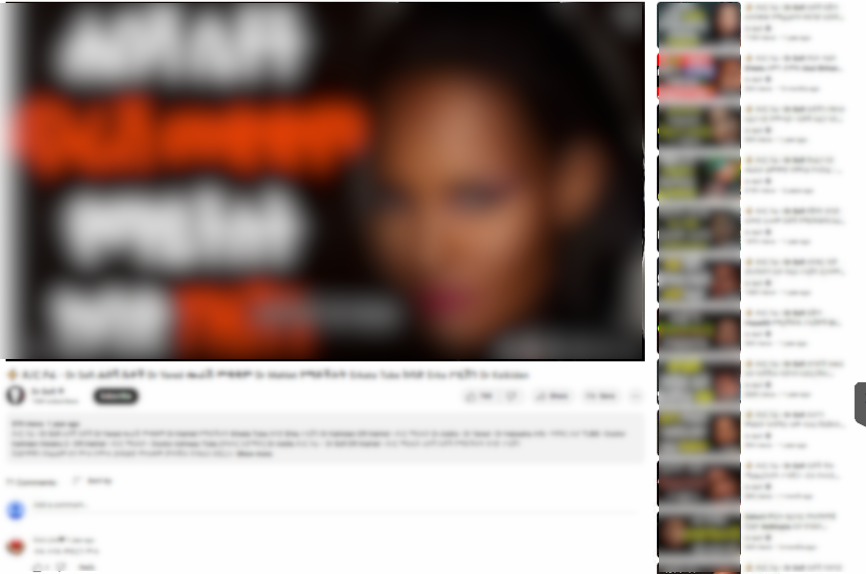}
        \caption{\textcolor{red}{\textbf{CW: Discussion of sexual content.}} Screenshot showing recommendations for one of the sexual videos opened to collect recommendation data. This video is from a verified channel that exclusively posts sexual videos in Amharic. All the recommendations for the one video opened from this channel are other sexual videos all from the same channel. The videos would have an image of a woman, often Ethiopian, and explicit sexual writing in Amharic. The channel also uses a `Dr.' title in their channel name.}
        \label{fig:verified}
    \end{figure}

\begin{figure}
        \centering
        \includegraphics[width=0.5\textwidth]{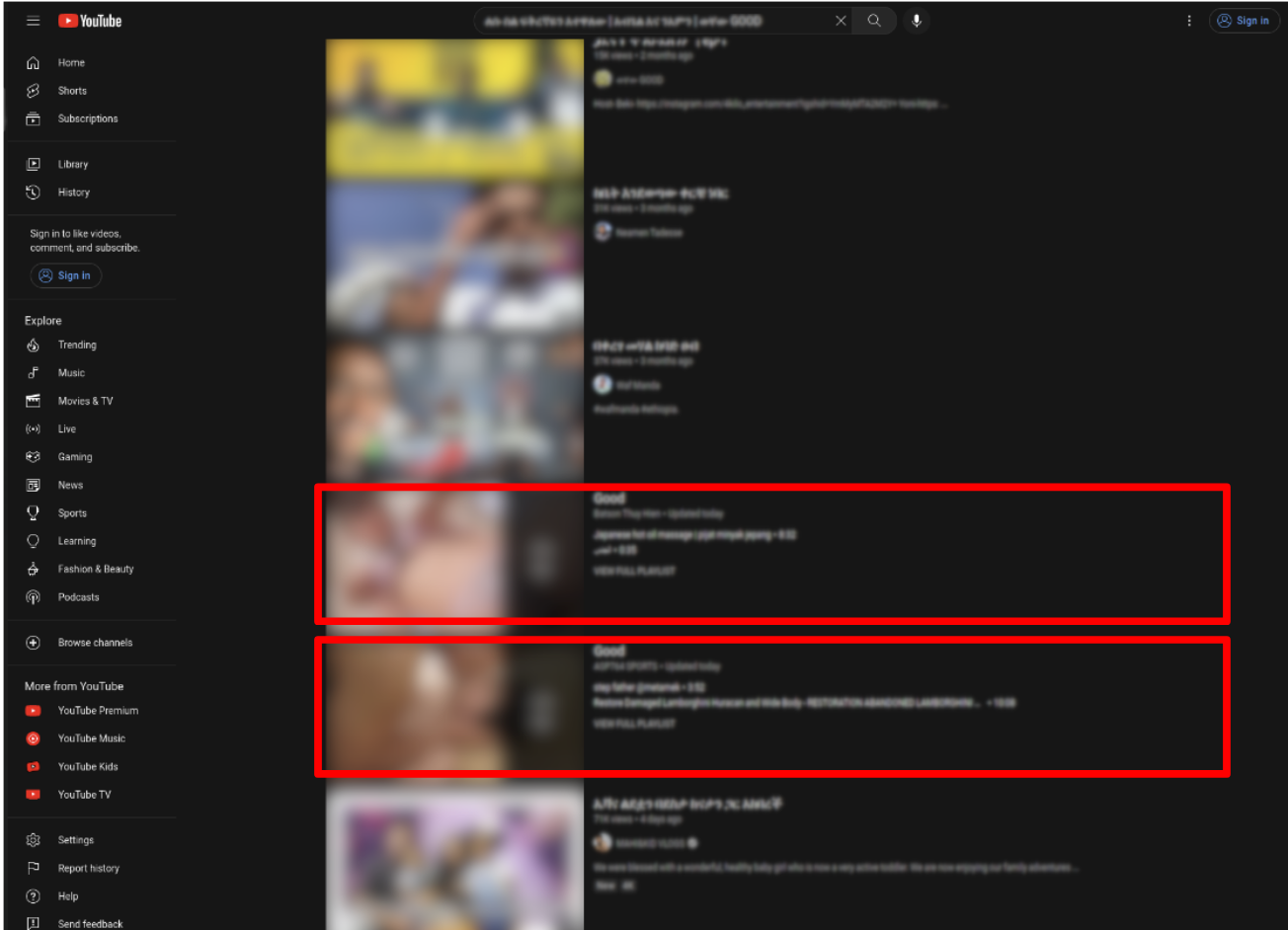}
        \caption{\textcolor{red}{\textbf{CW: Discussion of sexual content.}} In trying to label some videos from the collected data, we searched the titles of some videos on the YouTube interface. The results for the Ge'ez-based titles would sometimes return graphically explicit videos. In this case, the search query included the word "GOOD" which phonetically is similar to a word in Amharic used to describe astonishment. The fourth and fifth search results (highlighted in red boxes) were playlists with the title "Good" but had video thumbnails of actual pornographic videos.}
        \label{fig:graphicImage}
    \end{figure}

\begin{figure}
        \centering
        \includegraphics[width=0.4\textwidth]{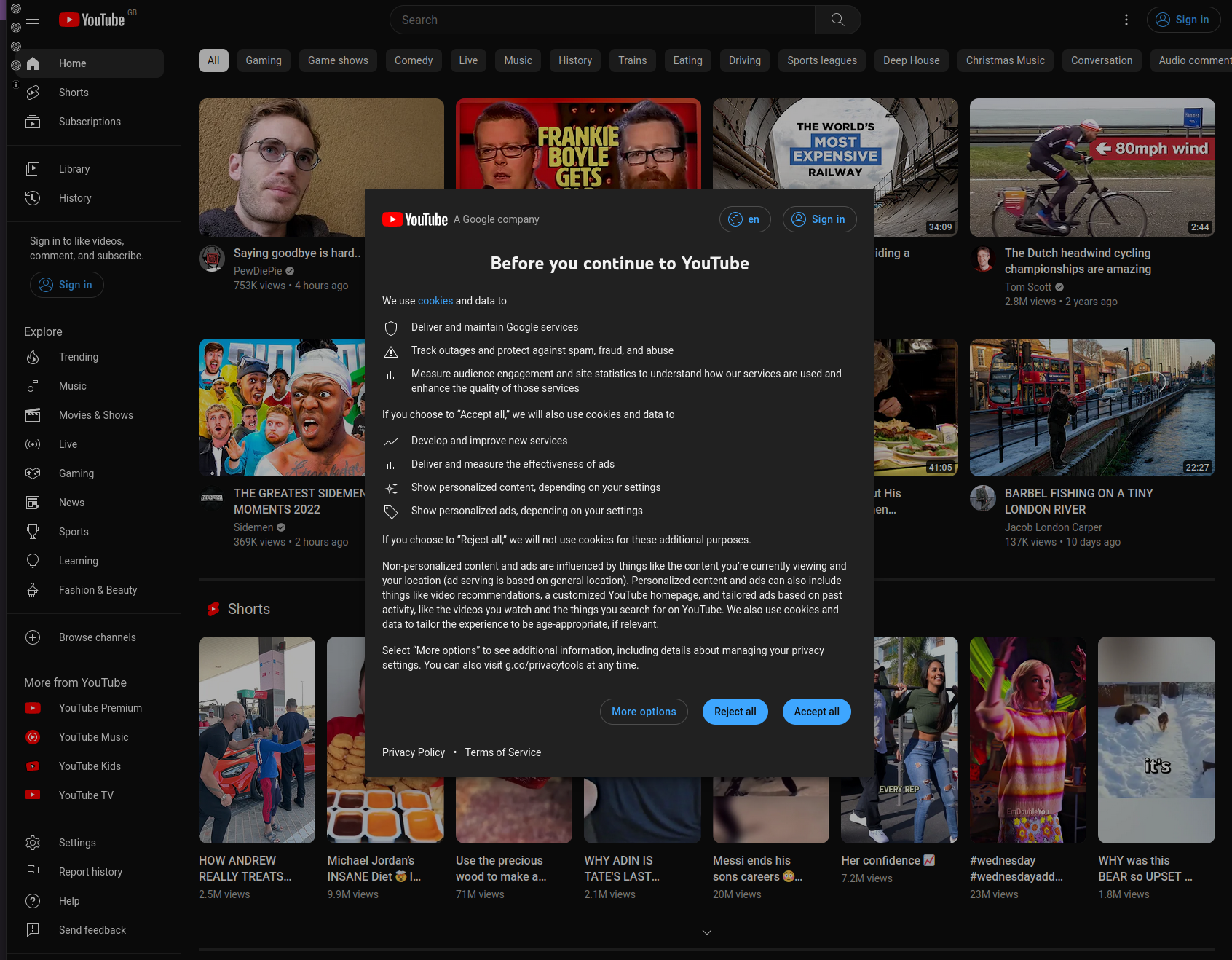}
        \caption{Screenshot showing privacy notice for opening YouTube when the VPN location was changed to the UK. The first thing on the YouTube interface was a prompt about data collection through cookies and options to accept all, reject all or ask for more options. YouTube did not display this notice for any of the other four locations.}
        \label{fig:privacy_UK}
    \end{figure}

In this section, we provide some screenshots from Study 1. In Fig. \ref{fig:search_benign}, we show how sexual videos occur in search results for benign queries. We found that sexual videos are not limited to just recommendation and search results; in Fig. \ref{fig:red}, when the query was the name of an Ethiopian celebrity, the ``People also watched'' section showed a video with a sexual thumbnail and title. As seen in Fig. \ref{fig:verified}, we also found that for a verified channel that posts sexual content in Amharic, the recommended videos were all sexual videos from the same channel, a clear violation of the policy to suppress content which does not abide by the community guidelines. Additionally, we noted that when we set our VPN location to the UK, we got a privacy notice about data collection and were offered options to limit the extent of data collection by YouTube (see Fig. \ref{fig:privacy_UK}). We did not receive this notice in any other location where we collected data from. During the labeling process, searching for video titles found in our collected data returned actual, non-Ethiopian pornographic videos (see Fig.\ref{fig:graphicImage}). In the recommendations, we observed a range of different types of videos. As seen in Fig. \ref{fig:rec_vids}, recommendations include, but are not limited to, sexual videos in other languages and sexual scenes cut from movies.

  \begin{figure}
          \centering
           \subfloat[Screenshot showing recommended videos once a sexual video is opened.]{\includegraphics[width=0.35\textwidth]{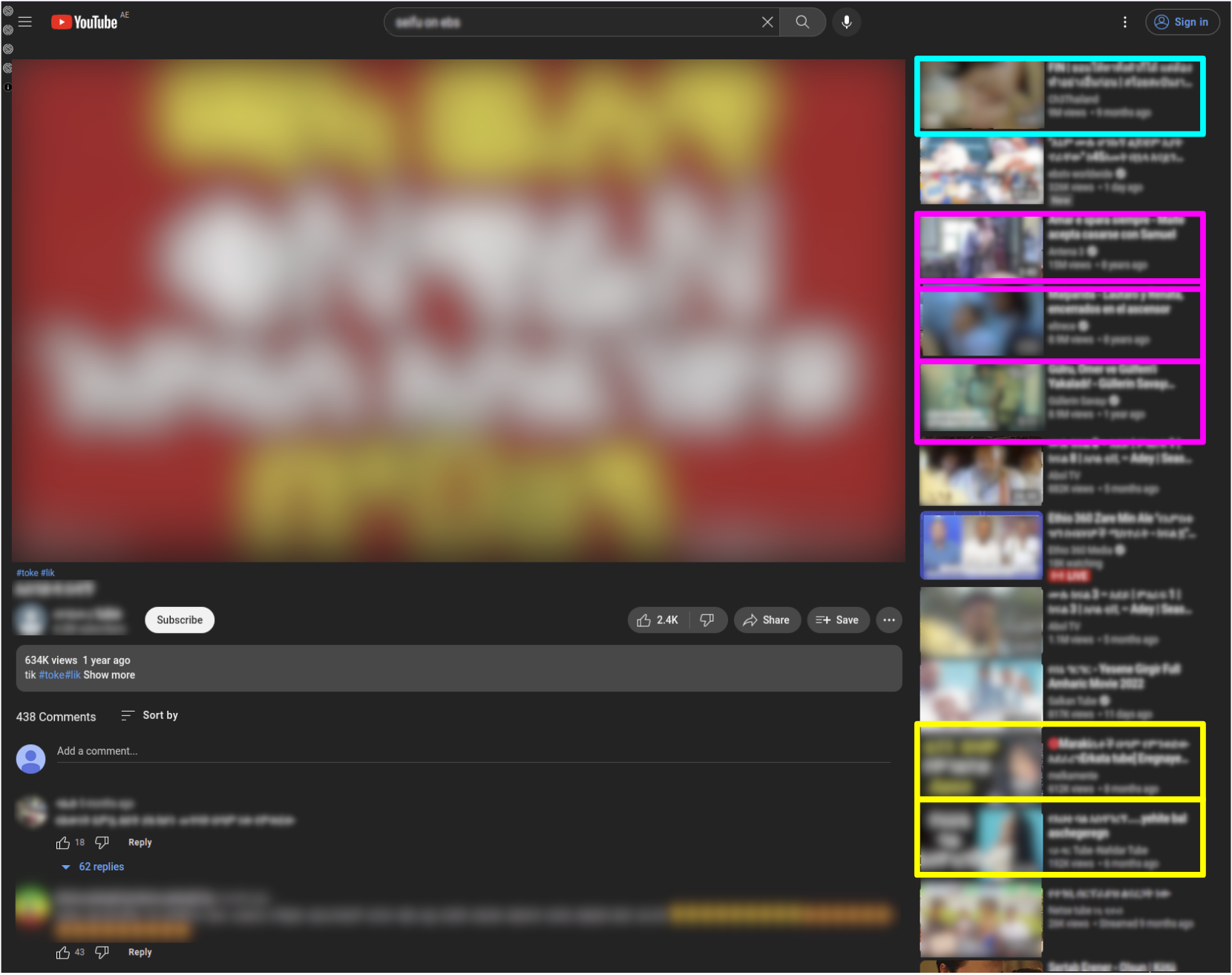}\label{fig:recs}}
          \hfill
          \subfloat[Screenshot showing recommended videos as one scroll through the recommendations.]{\includegraphics[width=0.35\textwidth]{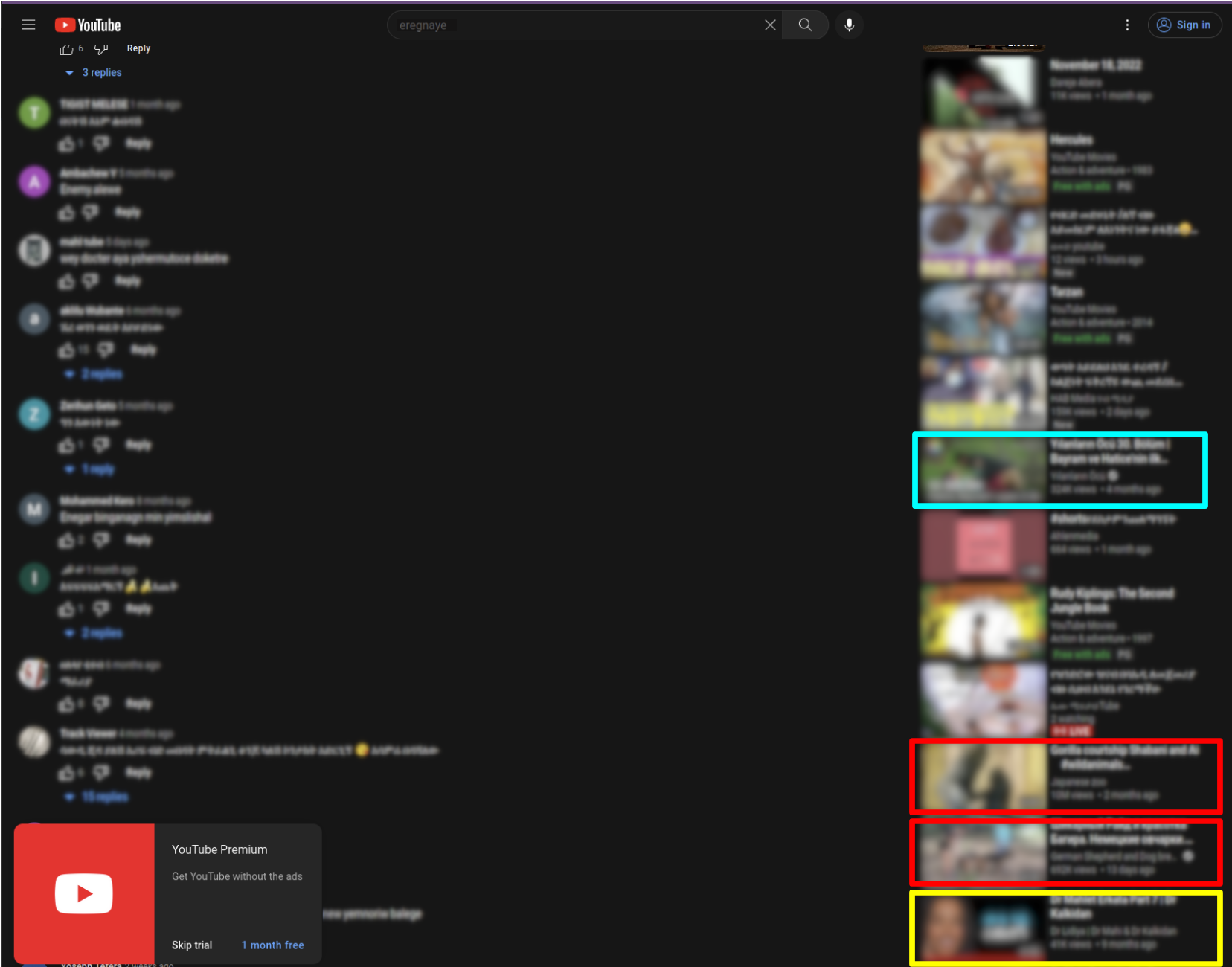}\label{fig:animals}}
            \hfill
          
          \caption{\textcolor{red}{\textbf{CW: Discussion of sexual content.}} Fig. \ref{fig:recs} shows a screenshot of the recommendation list for an Amharic sexual video we opened for data collection. The video has explicit Amharic writing and uses lexical variation by mixing Ge'ez and Latin characters. Fig. \ref{fig:animals} shows screenshots of recommended videos as we scroll down the recommendations for an Amharic sexual video. Recommendations include sexual videos in other languages (highlighted in \textcolor{teal}{teal}), sexual scenes cut from movies (highlighted in \textcolor{pink}{pink}), Amharic sexual videos from other channels (highlighted in \textcolor{yellow}{yellow}), and videos of animals mating (highlighted in \textcolor{red}{red}).}
          
      \label{fig:rec_vids}

        \end{figure}

    \begin{figure}
        \centering
        \includegraphics[width=0.4\textwidth]{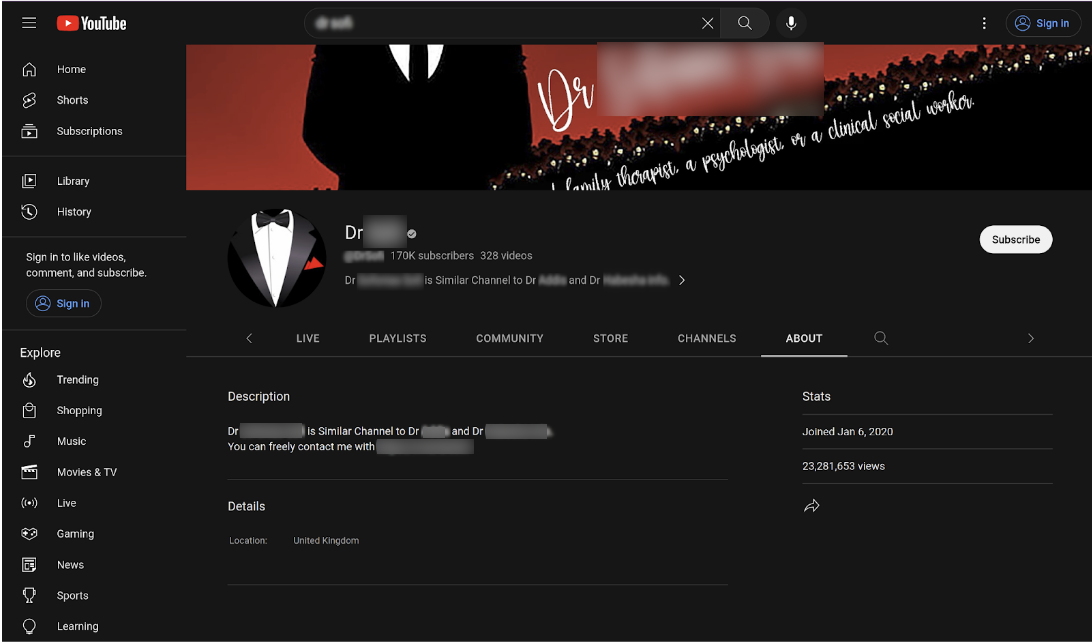}
        \caption{Screenshot showing a YouTube-verified channel that exclusively posts sexual videos in Amharic. The channel name (blurred to protect the identity of the creator) starts with ``Dr.'' followed by a popular Ethiopian name. Further in the description, the channel states ``Dr. [NAME\_1] is similar channel to Dr. [NAME\_2] and Dr. [NAME\_3]. You can freely contact me with [TELEGRAM LINK]''. The channel location is set in the United Kingdom and has over 23 million views.}
        \label{fig:doc}
    \end{figure}

\onecolumn
\section{Additional Evidence for Study 2} \label{apen:evidence}

\begin{figure}[h]
    \centering
    \includegraphics[width=0.8\textwidth]{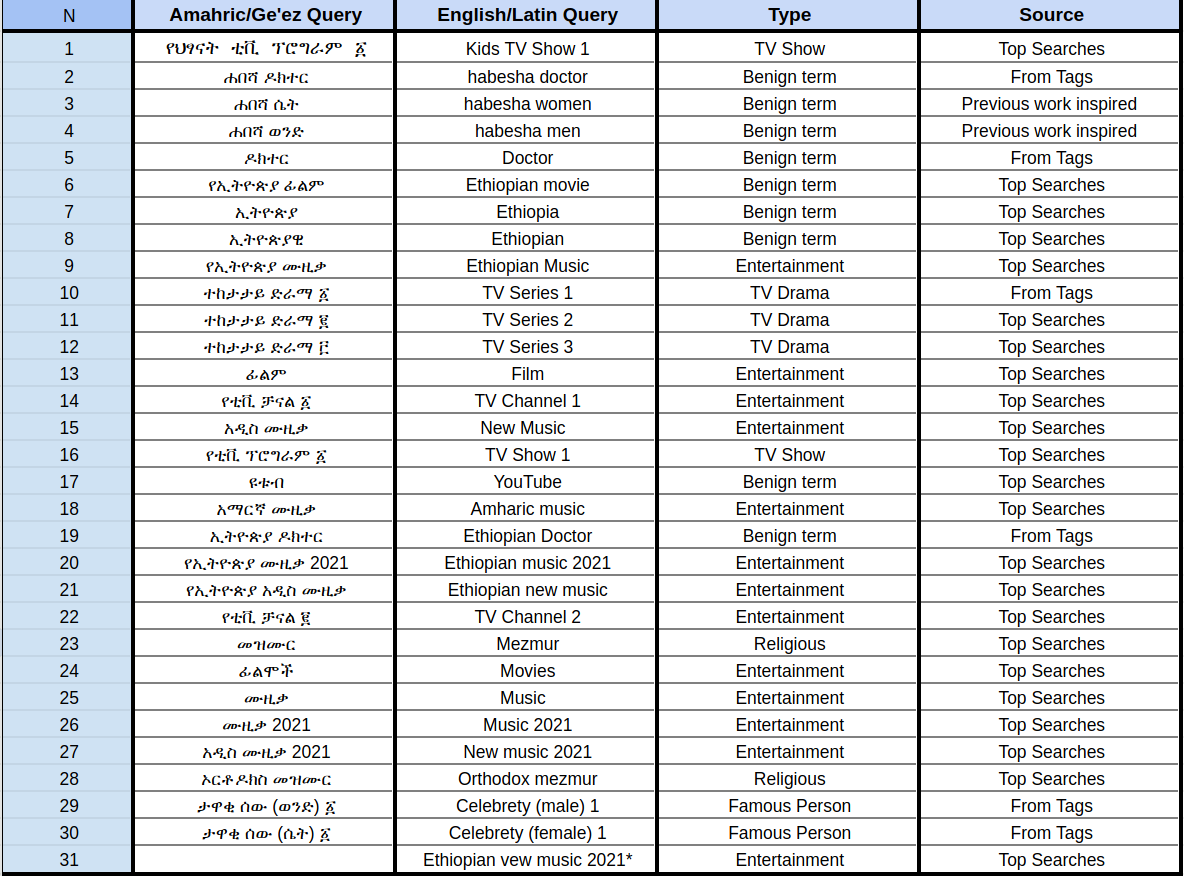}
    \caption{\textbf{Search queries we curated to collect data for \hyperref[study1]{Study 1}.} In this table, we show the Ge'ez and the Latin versions of each query which were all used in our data collection. We curated the search queries based on (1) top YouTube search statistics for Ethiopia, (2) popular, benign terms from sexual videos we were getting during our initial data collection, and (3) benign phrases inspired by previous work that studies search engine bias. The last query on the Latin list does not have an Amharic equivalent since it includes the misspelled word `vew' in place of `new'; we still ran the query in English as it was in the Top YouTube searches. Following guidelines in \textsc{HarmCheck}, we replaced celebrity names with identifiers, to avoid exposing their identity and any potential association with harm.}
    \label{fig:queries}\vspace*{-12pt}
\end{figure}

\begin{figure}[h]
          \centering
           \subfloat[Distribution of Sexual videos per query for Ge'ez based search.]{\includegraphics[width=0.5\textwidth]{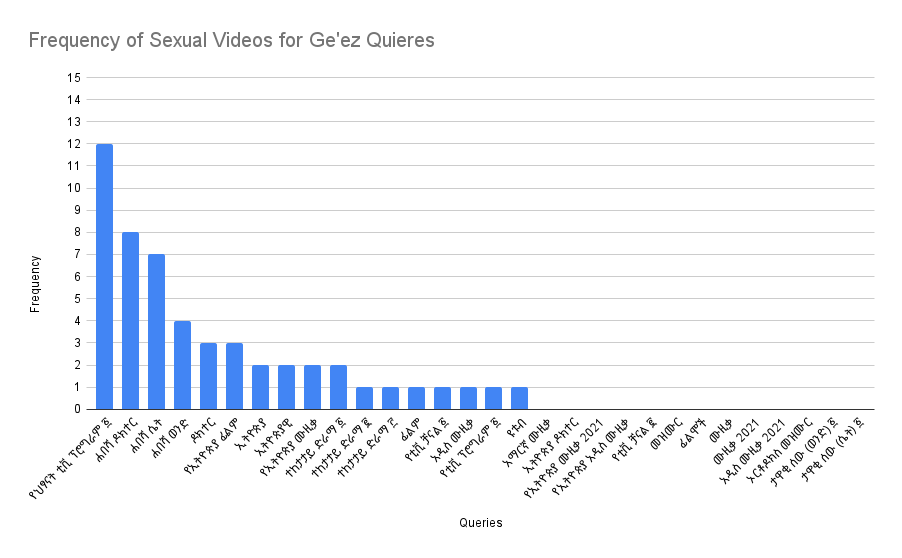}\label{fig:geez_dist}}
          \hfill
          \subfloat[Distribution of Sexual videos per query for Latin-based search.]{\includegraphics[width=0.5\textwidth]{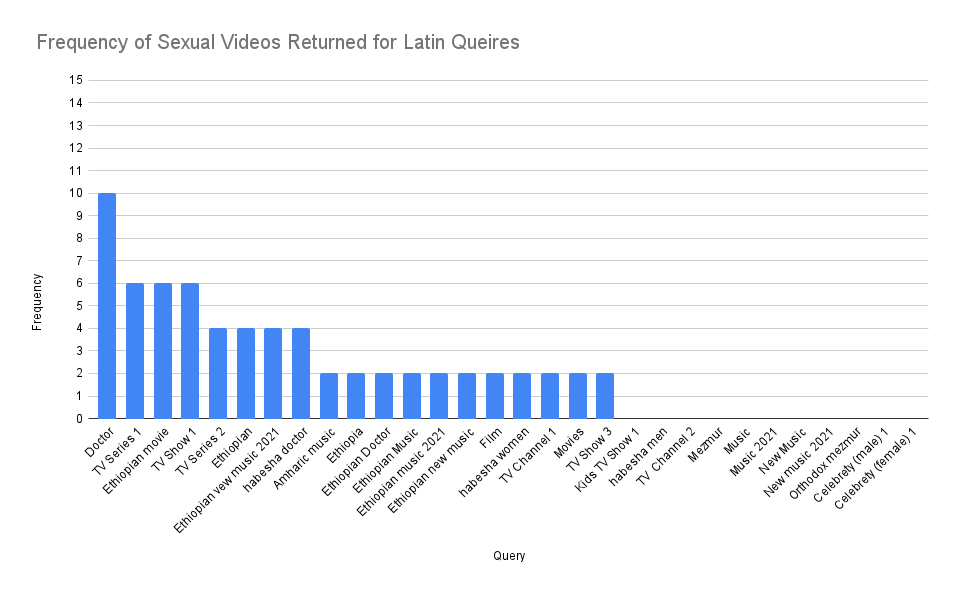}\label{fig:latin_dist}}
            \hfill
          
          \caption{Fig. \ref{fig:geez_dist} shows the distribution of sexual videos for each of the Ge'ez queries in our Study 2. We see that the highest number of sexual videos is returned for a query with a children's show name. Fig. \ref{fig:latin_dist} shows the distribution for the Latin queries in Study 2; we see that the highest number of sexual videos is returned for the query ``Doctor''. }
          
      \label{fig:rec_vids}
      \vspace*{-6pt}
      
        \end{figure}

In this section, we provide further evidence to support our findings in Study 2. In Fig. \ref{fig:queries}, we present the full list of queries we assembled as described in Sec. \ref{study2}. In Fig. \ref{fig:geez_dist} and Fig. \ref{fig:latin_dist}, we show the distribution of policy-violating sexual content returned for the queries we assembled for Study 2. In Table \ref{tab:recdata}, we show the percentage of each of the categories of policy-violating videos in recommendations. 

    \begin{table*}
          
          \begin{tabular}{p{1.7cm}|p{7.5cm}|ccccc}
            \toprule
           \textbf{Video Label} & \textbf{Label Description} &\textbf{Ethiopia}&\textbf{Saudi Arabia} & \textbf{UAE} & \textbf{US} & \textbf{UK}\\
            \midrule
            Sexual & Narrations of sexual stories; audio recordings of sexual intercourse and phone sex; expos\'e videos with fully or partially naked people performing sexual acts; sexual videos in other, non-English languages; sex scenes cut from movies; videos of animals mating; rape advocating videos; documentation of rape acts; videos with sexually explicit thumbnails but educational content; non-consensual released sexual videos. & \textbf{45.15\%} & 15.78\% & \textbf{35.52\%} & 18.67\% & \textbf{38.25\%}\\
            \hline
            Entertainment & Talk shows; games and competitions; and food channels & 30.90\% & \textbf{33.73\%} & 33.16 & \textbf{34.09\%} & 30.96\%\\
            \hline
            Politics & News; Commentary; Interviews & 6.36\% & 13.59\% & 4.04\% & 3.43\% & 9.65\%\\
            \hline
            Vlogs & Video blogs by college students; Ethiopian people living in the Middle East & 0.30\% & 12.70\% & 8.08\% & 5.33\% & 11.84\%\\
            \hline
            Educational & videos about taxes; high school textbook explanations; and a decent amount of language learning videos & 8.18\%  & 8.92\% & 5.38\% & 14.48\% & 0.91\%\\
            \hline
            Religious & Songs and sermons from Ethiopian Orthodox Christian, Islam, and Evangelical Christian religions. & 3.63\% & 7.34\% & 5.55\% & 11.81\% & 4.01\%\\
            \hline
            Sports & News; Commentary; Recorded Streams; Reaction Videos. & 0.45\% & 1.49\% & 1.34\% & 4.04\% & 0.18\%\\
            \hline
            Motivational & Talks; Interviews; Narrations & 2.12\% & 2.28\% & 2.69\% & 5.14\% & 1.82\% \\
            \hline
            Relationship Advice & Podcasts; Interviews; Video essays & 2.42\% & 1.69\% & 3.53\% & 2.29\% & 1.46\%\\
            \hline
            Information & Where to buy things; Reviews & 0.45\%  & 2.48\% & 0.17\% & 0.57\% & 0.00\%\\
            \hline
            Conspiracy Theories  & Interviews; Video essays & 0.00\% & 0.00\% & 0.50\% & 0.19\% & 0.91\%\\
          \bottomrule
            
        \end{tabular}
          \caption{\textcolor{red}{\textbf{CW: Discussion of sexual content and rape.}}\textbf{Types of recommended videos for every policy-violating sexual video identified, by country.} For each of the policy-violating sexual videos we identified, we categorize the videos recommended. We observed that for Ethiopia, UAE, and UK locations, clicking on policy-violating sexual content will likely lead to more sexual content being recommended by YouTube's algorithm. Of all locations, Ethiopia has the highest recommendation for sexual videos, with 45.15\% of recommendations being sexual videos.}\label{tab:recdata}      \vspace*{-6pt}

    \end{table*}

\clearpage
\section{Themes from Interview Study} \label{apn:themes}
In this section, we provide themes and sub-themes that emerged from our interview study analysis and provide example open codes in Table \ref{tab:themes}. 
\begin{table*}[b]
    \centering
    \begin{tabular}{p{3cm}|p{5.5cm}|p{6cm}}
    \hline
        \textbf{Second-Level Themes} &\textbf{ First-Level Themes} & \textbf{Example open codes}\\
        \hline
        \multirow{3}{3cm}{Users' search experience} & experience based on search language & report better experience searching for English content \\
        & experience based on writing script & indicate searching in Ge'ez is harder due to keyboard switching\\
        & mixed, foreign languages in search results &  got content in Hindi for Amharic query \\
        \hline
        \multirow{5}{3cm}{Perceptions of how YouTube features and procedures work} & perceptions about video recommendations &  believes recommendations are usually based on watch history\\
        & perceptions about Home Page &  get political content recommendation on home page \\
        &perceptions about trends & trend has more or less evolved with their taste\\
        &perceptions of effects of VPN & search experience are the same with and without VPN\\
        &perceptions about YouTube policy & believes explicit language is not allowed \\
        &perceptions about reporting mechanisms & believes there is a queue for reports\\
         \hline
        \multirow{3}{3cm}{Policy, culture, and fairness} & questions on who makes decisions & questioning the cultural context `sexual videos' are defined in \\ 
        &role of culture in justification of content & notes videos use conservative culture to justify releasing degrading, non-consensual videos of women \\
        & (anticipated) consequence of unaddressed policy-violating content & fears unintended exposure of younger generation to sexual content\\ 
         \hline
        \multirow{7}{3cm}{YouTube platform usage and utility} & default YouTube settings & default YouTube language is English \\
        &time span of use & has been using YouTube for over 8 years\\
        &comment feature & commented to defend a woman whose video was released without her consent\\
        &like feature & likes to get the `algorithm' to bring them similar content\\
        &posting videos & has never posted their own video\\ 
        &subscriptions &  subscribes to channels after checking the full video list \\
        &substitute platform & does not believe there is a substitute for YouTube \\
        
         \hline
       \multirow{4}{3cm}{ Harmful content} &  characters of sexual videos & has been exposed to non-consensual release of sexual video if a minor \\
       &categories of other harmful content & has been exposed to graphic and violent content\\
       &events that led to unintended exposure of sexual videos &  got exposed while searching for a famous singer\\
       &strategies against unintended exposure & uses multiple google accounts \\
         \hline
        \multirow{3}{3cm}{Reporting practices} & steps taken to report & chose sexually explicit from the provided categories\\
         &outcomes of reporting & did not receive feedback on their email\\
         &types of reported videos & ethnic hate speech video\\
         \hline
    \end{tabular}
    \caption{\textbf{Second- and first-level themes, and example open codes from Study 2.} Our analysis resulted in 936 unique open codes which were grouped to 26 first-level themes, which were further grouped into 6 second-level themes.}
    \label{tab:themes}
\end{table*}

\newpage

\end{document}